\documentclass[12pt,a4paper]{article}
\pdfoutput=1
\usepackage[utf8]{inputenc}
\newcommand{\sphid}[1]{}

\usepackage{epsf,amsmath,amssymb,graphicx,dcolumn}
\usepackage{caption}
\usepackage[labelformat=simple]{subcaption}

\usepackage{scalefnt,ulem,pstricks}
\usepackage{booktabs,multirow,tabularx}
\usepackage[colorlinks=true,allcolors={blue!70!black}]{hyperref}
\usepackage{cleveref}
\usepackage{color}
\usepackage{rotating}
\usepackage{microtype}
\usepackage[titletoc,title]{appendix}
\usepackage[numbers,sort&compress]{natbib}
\usepackage{amsmath,amsfonts,amsthm,bm}
\usepackage{xspace}
\providecommand{\href}[2]{#2}

\newcommand{\tmop}[1]{\ensuremath{\operatorname{#1}}}

\newcommand\F{$Q\bar Q$}
\newcommand\FJ{$Q\bar Q{\rm J}$}
\newcommand\FJJ{$Q\bar Q{\rm JJ}$}

\newcommand{\as}{\alpha_s}

\newcommand{\pt}{{p_{\text{\scalefont{0.77}T}}}}

\newcommand{\ptrad}{{p_{\text{\scalefont{0.77}T,rad}}}}

\newcommand{\muF}{{\mu_{\text{\scalefont{0.77}F}}}}
\newcommand{\muR}{{\mu_{\text{\scalefont{0.77}R}}}}

\newcommand{\muRc}{{\mu_{\text{\scalefont{0.77}R}}^{(0)}}}
\newcommand{\KF}{{K_{\text{\scalefont{0.77}F}}}}
\newcommand{\KR}{{K_{\text{\scalefont{0.77}R}}}}

\newcommand{\noun}[1]{{\scshape #1}}

\newcommand{\POWHEG}{\noun{Powheg}}

\newcommand{\POWHEGBOXRES}{\noun{Powheg-Box-Res}}

\newcommand{\minlo}{{\noun{MiNLO$^{\prime}$}}\xspace}
\newcommand{\minnlo}{{\noun{MiNNLO$_{\rm PS}$}}\xspace}

\newcommand{\OpenLoops}{{\noun{OpenLoops}}\xspace}

\newcommand{\PYTHIA}[1]{\noun{Pythia{#1}}\xspace}

\newcommand{\citere}[1]{Ref.\,\cite{#1}}

\newcommand{\citeres}[1]{Refs.\,\cite{#1}}

\newcommand{\eqn}[1]{Eq.\,(\ref{#1})}

\newcommand{\fig}[1]{Figure\,\ref{#1}}

\newcommand{\tab}[1]{Table\,\ref{#1}}
\newcommand{\sct}[1]{Section~\ref{#1}}

\newcommand{\app}[1]{Appendix~\ref{#1}}
\newcommand{\LambdaPWG}{\Lambda_{\rm pwg}}

\usepackage{etoolbox}
\makeatletter
\patchcmd{\@sect}{#8}{\boldmath #8}{}{}
\let\ori@chapter\@chapter
\def\@chapter[#1]#2{\ori@chapter[\boldmath#1]{\boldmath#2}}
\makeatother

\usepackage{scalerel}

\newcommand{\MSbar}{\ensuremath{\overline{\text{MS}}}}

\usepackage[tikz]{bclogo}
\usepackage{subcaption}
\usepackage{tikz-feynman}
\usepackage{cancel}
\usepackage{verbatim}
\usetikzlibrary{arrows,shapes}
\usepackage{afterpage}

\begin{document} 
\begin{flushright}
\vspace*{-1.5cm}
MPP-2026-180 
\end{flushright}
\vspace{0.cm}

\begin{center}
{\Large \bf The production of bottom-flavoured jets at the LHC \\[0.2cm] 
through bottom-quark pair production at NNLO+PS}
\end{center}

\begin{center}
{\bf Rhorry Gauld$^{(a)}$}, {\bf Alessandro Ratti$^{(a,b)}$}, \\[1ex]
{\bf Marius Wiesemann$^{(a,c)}$}, and {\bf Giulia Zanderighi$^{(a,b)}$}

(a) Max-Planck-Institut f\"ur Physik, Boltzmannstraße 8, 85748 Garching, Germany

(b) Technische Universit\"at M\"unchen, James-Franck-Strasse 1, 85748 Garching, Germany

(c) Fakult\"at Physik, Technische Universit\"at Dortmund, 44227 Dortmund, Germany

\href{mailto:rgauld@mpp.mpg.de}{\tt rgauld@mpp.mpg.de}\\
\href{mailto:ratti@mpp.mpg.de}{\tt ratti@mpp.mpg.de}\\
\href{mailto:marius.wiesemann@mpp.mpg.de}{\tt marius.wiesemann@mpp.mpg.de}\\
\href{mailto:zanderi@mpp.mpg.de}{\tt zanderi@mpp.mpg.de}

\end{center}

\begin{center} {\bf Abstract} \end{center}\vspace{-1cm}
\begin{quote}
  \pretolerance 10000
  
We consider the dominant production
mechanism of bottom-flavoured jets at the LHC, which proceeds through bottom-quark pair production. We compute next-to-next-to-leading-order (NNLO) QCD corrections
retaining full bottom-quark mass effects and match them with a parton shower
(NNLO+PS). Our predictions provide a realistic description of events with one
and two bottom-flavoured jets at both the partonic and hadronic level. We
compare different jet-flavour definitions and confirm that the
dominant numerical differences between the jet algorithms result from the treatment of jets that contain two bottom quarks originating from $g\to b\bar{b}$
splittings. We compare our predictions, using the experimental bottom-flavour tagging prescription, with measurements by ATLAS, CMS, and LHCb at 7 and 13\,TeV, finding good agreement for the distributions presented by ATLAS and CMS and a good description of the shapes of the LHCb data.
We also discuss the impact of various physics effects on the considered distributions: power corrections in the bottom-quark mass, the implications of symmetric/asymmetric jet-selection cuts, as well as contributions from multi-parton interactions.

\end{quote}

\parskip = 1.2ex

\newpage
\setcounter{tocdepth}{1}
\tableofcontents

\section{Introduction}

The rich physics programme at the Large Hadron Collider (LHC) has reached 
the precision frontier. The significant pace of the data taking by the experiments in the High-Luminosity LHC (HL-LHC) runs  
will further decrease the experimental uncertainties in the 
coming years, challenging the precision of many of today's 
theoretical predictions. In fact, precision tests of the Standard Model (SM) through
data--theory comparisons have become one of the main avenues for probing 
new-physics phenomena during the LHC era. In essentially every LHC measurement, the reconstruction of jets plays a central role, both in defining the signal and in suppressing backgrounds. More generally, jets provide a simplified description of the complex hadronic final state, enabling the reconstruction of the underlying hard scattering process, or the decay topology of heavy particles. For example, jet substructure can be exploited to discriminate between two-prong and three-prong decays.
Another important characteristic of the hard process is the flavour of the jets, namely whether they originate from light quarks or gluons, or instead from heavy quarks such as charm or bottom quarks.

Processes with bottom-flavoured (or charm-flavoured) jets in the final 
state play an increasingly 
important role at the LHC. For instance, they may be produced in association with 
other particles, such as a Higgs or a vector boson, or they can be produced in decays
of heavy particles, like a Higgs boson or a top quark. The dominant production mechanism
of (one or more) bottom-flavoured jets ($b$-jets), however, is the QCD production of a bottom-quark
pair in the hadronic collision.
This process is both an irreducible background to many signal processes, such as those mentioned above, and a valuable benchmark for precision QCD studies.
In particular, it does constitute one of the most abundant signatures to directly
study bottom-flavoured jets at the LHC. 
The topic of jet-flavour assignment has also received considerable theoretical interest in recent years~\cite{Buckley:2015gua,Goncalves:2015prv,Caletti:2022hnc,Caletti:2022glq,Czakon:2022wam,Gauld:2022lem,Caola:2023wpj,Larkoski:2023upz,Larkoski:2024nub,Behring:2025ilo,Larkoski:2025afg,Generet:2025gdy}, which has now been considered in experimental measurements through unfolding procedures \cite{ATLAS:2024tnr}, and may have implications for future measurements of processes involving flavoured jets. 
An important open question is how the suggested approaches can be reconciled with the jet-flavour tagging by the experiments.

In this respect, one-$b$-jet or two-$b$-jet-based measurements, where the
signal $b$-jets predominantly originate from bottom-quark ($b\bar{b}$) pair production in the 
hard process, can provide particularly useful insights to the definition of the 
(bottom) flavour of jets. These processes have been measured at the LHC and in the past
by various experiments: First measurements of $B$-hadron production in 
proton--anti-proton collisions were already 
performed at CERN's Super Proton Synchrotron (SPS) by the UA1 collaboration
\cite{UA1:1990vvp}. Later also $b$-jet studies have been performed at the Fermilab's Tevatron by CDF \cite{CDF:2008hmn} and D0 \cite{D0:2000vvk}. At the LHC,
all four experiments, including ALICE \cite{ALICE:2021wct}, ATLAS \cite{ATLAS:2011ac,ATLAS:2016anw,ATLAS:2018zhf}, CMS \cite{CMS:2012pgw}, and LHCb \cite{LHCb:2014jms,LHCb:2017cyt,LHCb:2020frr,LHCb:2025tvf},
have presented measurements of $b$-jet cross sections in proton--proton collisions 
at various center-of-mass energies.

On the theoretical side, considerable effort has been devoted to improving the accuracy of predictions for processes involving heavy-flavour production.
At typical LHC energies, bottom quarks can be described either
as massless partons in the five-flavour scheme (5FS) or as massive quarks in
the four-flavour scheme (4FS). Bottom-quark pair
production has been studied extensively at fixed order, including NLO
calculations, in some cases including a resummation of logarithmically enhanced contributions~\cite{Nason:1987xz,Nason:1989zy,Beenakker:1988bq,Mangano:1991jk,
Cacciari:1993mq,Cacciari:1998it,Cacciari:2001cw,
Cacciari:2001td,Cacciari:2002pa,Kniehl:2005mk,
Kramer:2018vde,Benzke:2019usl,Cacciari:2012ny,Aliev:2010zk} and
more recently NNLO predictions~\cite{Catani:2020kkl,Czakon:2024tjr}.
When considering $b$-jets there are subtle infrared properties of
their definitions, which complicate fixed-order calculations with massless
bottom quarks, and the sensitivity of $b$-jets to effects beyond fixed-order
matrix elements, such as parton showers, hadronization, and multiparton
interactions. 

Due to the relatively low characteristic scale of the process, the strong
coupling is sizable, leading to sizeable higher-order QCD corrections.
Dedicated fixed-order predictions for inclusive $b$-jet production have been
obtained at NLO only more recently~\cite{Bierenbaum:2016gzb}, while first NNLO results for $b$-jet production in the massless scheme have been presented in \citere{Generet:2025gdy}.
Electroweak corrections to $b$-jet production in the massless~\cite{Kuhn:2009nf} and massive schemes~\cite{Gauld:2019doc} have also been presented in various contexts.
Heavy-quark fragmentation and jet-fragmentation approaches have further allowed the
resummation of logarithms associated with the bottom-quark mass and the jet
radius~\cite{Dai:2018ywt,Li:2018xuv,Czakon:2024tjr}. Realistic
phenomenological $b$-jet predictions are therefore obtained by combing 
higher-order calculations with parton shower, with various NLO implementations available in
different schemes and frameworks~\cite{Frixione:2007nw,Buonocore:2017lry,
Alwall:2014hca,Sherpa:2019gpd,Sherpa:2024mfk}.

In this article we study $b$-jet production in hadronic collisions
at next-to-next-to-leading order (NNLO) in QCD retaining finite bottom-mass effects. 
In addition to the
NNLO QCD corrections to the $b\bar{b}$ core process we include  
effects from the parton showering that are relevant to appropriately 
model $b$-jets. The results are based on the NNLO+PS calculation for $b\bar{b}$ production presented in \citere{Mazzitelli:2023znt}.
In this context, we study heavy-flavour jet-definitions, that rely on
different procedures to assign the flavour of a jet.
We exploit our results to compare our predictions with measurements by
different LHC experiments at $7$ and $13$\,TeV, finding overall very good
agreement with the data. Finally, we investigate the role of finite
bottom-quark mass effects by extracting the corresponding power
corrections and comparing massive and massless predictions.

\section{Outline of the calculation}

\begin{figure}[t]
  \begin{center}
    \begin{subfigure}[b]{.3\linewidth}
      \centering
\begin{tikzpicture}
\begin{feynman}
	\vertex (a1) at (0,0) {\( q\)};
	\vertex (a2) at (0,-2) {\(\bar q\)};
	\vertex (a3) at (1.4,-1);
	\vertex (a4) at (2.7,-1);
	\vertex (a5) at (4.1,0){\( b\)};
	\vertex (a6) at (4.1,-2){\(\bar b\)};
        \diagram* {
          {[edges=fermion]
            (a1)--[thick](a3)--[thick](a2),
            (a6)--[fermion, ultra thick](a4)--[fermion, ultra thick](a5),
          },
          (a3) -- [gluon,thick] (a4),
        };
      \end{feynman}
\end{tikzpicture}
\caption{$s$-channel $q\bar{q}$ diagram}
        \label{subfig:qq}
\end{subfigure}%
\begin{subfigure}[b]{.3\linewidth}
  \centering
\begin{tikzpicture}
  \begin{feynman}
	\vertex (a1) at (0,0) {\( g\)};
	\vertex (a2) at (0,-2) {\( g\)};
	\vertex (a3) at (1.4,-1);
	\vertex (a4) at (2.7,-1);
	\vertex (a5) at (4.1,0){\( b\)};
	\vertex (a6) at (4.1,-2){\(\bar b\)};
        \diagram* {
          {[edges=fermion]
            (a6)--[fermion, ultra thick](a4)--[fermion, ultra thick](a5),
          },
          (a3) -- [gluon,thick] (a4),
          (a2)--[gluon,thick](a3)--[gluon,thick](a1),
        };
  \end{feynman}
\end{tikzpicture}
\caption{$s$-channel $gg$ diagram}
        \label{subfig:gg}
\end{subfigure}%
\begin{subfigure}[b]{.3\linewidth}
  \centering
\begin{tikzpicture}
  \begin{feynman}
	\vertex (a1) at (0,0) {\( g\)};
	\vertex (a2) at (0,-1.7) {\( g\)};
	\vertex (a3) at (1.53,0);
	\vertex (a4) at (1.53,-1.7);
	\vertex (a5) at (3,0){\( b\)};
	\vertex (a6) at (3,-1.7){\(\bar b\)};
        \diagram* {
          {[edges=fermion]
            (a6)--[fermion, ultra thick](a4)--[fermion, ultra thick](a3)--[fermion, ultra thick](a5),
          },
          (a2)--[gluon,thick](a4),
          (a3)--[gluon,thick](a1),
        };
  \end{feynman}
\end{tikzpicture}\vspace{0.15cm}
\caption{$t$-channel $gg$ diagram}
        \label{subfig:gg}
\end{subfigure}
\end{center}
\caption{\label{fig:diagrams} Feynman diagrams for 
  the QCD process $pp\to b\bar{b}$ at LO.}
\end{figure}
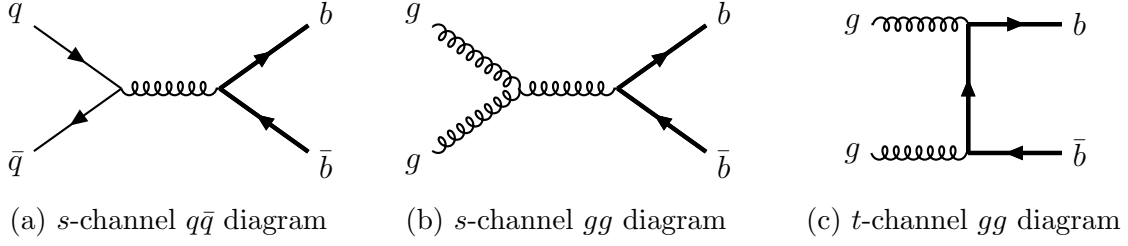

Our simulation of $b$-jet production at the LHC is based on the NNLO+PS
generator for bottom-quark pair production developed in \citere{Mazzitelli:2023znt}, which was 
achieved by following the \minnlo{} method for a heavy-quark pair \cite{Mazzitelli:2020jio,Mazzitelli:2021mmm}.
A selection of leading-order (LO) Feynman diagrams to the core process are shown in \fig{fig:diagrams}. 
The \minnlo{} method was initially
developed for colour-singlet production~\cite{Monni:2019whf,Monni:2020nks}.
Its extension to heavy-quark pair (\F{}) 
production \cite{Mazzitelli:2020jio,Mazzitelli:2021mmm} and the associated production of a heavy quark pair with colour singlets (\F{}$F$) \cite{Mazzitelli:2024ura}
were major breakthroughs and \minnlo{} is still the only NNLO+PS method that 
has been applied to processes with colour charges in the initial and the final state.
These developments lead to the construction of \minnlo{}
simulations for a broad range of LHC
processes~\cite{Lombardi:2020wju,Lombardi:2021rvg,Mazzitelli:2020jio,Mazzitelli:2021mmm,Buonocore:2021fnj,Lombardi:2021wug,Zanoli:2021iyp,Gavardi:2022ixt,Haisch:2022nwz,Lindert:2022qdd,Mazzitelli:2023znt,Mazzitelli:2024ura,Biello:2024vdh,Niggetiedt:2024nmp,Biello:2024pgo,Biello:2026nhj,Garosi:2026jyo}.

The \minnlo{} master formula is expressed in terms of a \POWHEG{} \cite{Nason:2004rx,Nason:2006hfa,Frixione:2007vw,Alioli:2010xd} NLO+PS 
calculation for the production of a heavy quark pair in association with a jet (\FJ{}):\footnote{We refer the interested reader to \citere{Mazzitelli:2021mmm} for the complete description and detailed derivation of the method. For the sake of brevity, we adopt a rather simplified and symbolic notation here.}
\begin{align}
\label{eq:master}
{\rm d}\sigma_{\scriptscriptstyle Q\bar Q}^{\rm MiNNLO_{PS}}={\rm d}\Phi_{\scriptscriptstyle Q\bar Q{\rm J}}\,\bar{B}^{\,\rm MiNNLO_{\rm PS}}\,\times\,\left\{\Delta_{\rm pwg}(\Lambda_{\rm pwg})+ {\rm d}\Phi_{\rm rad}\Delta_{\rm pwg}(p_{T,{\rm rad}})\,\frac{R_{\scriptscriptstyle Q\bar Q{\rm J}}}{B_{\scriptscriptstyle Q\bar Q{\rm J}}}\right\}\,,
\end{align}
where the \POWHEG{} $\bar B$ function is modified in such a way that it reaches NNLO QCD
accuracy for \F{} production when the additional jet becomes unresolved
\begin{align}
\label{eq:minnlo}
\bar{B}^{\,\rm MiNNLO_{\rm PS}}\sim \sum_{i=1}^{n_c}\mathcal{C}_i\,e^{-S_i}\,\bigg\{{\rm d}\sigma^{(1)}_{\scriptscriptstyle Q\bar Q{\rm J}}\big(1+S_i^{(1)}\big)+{\rm d}\sigma^{(2)}_{\scriptscriptstyle Q\bar Q{\rm J}}+D_i^{(\ge 3)}\times F^{\rm corr}\bigg\}\,.
\end{align}
With $\Phi_{\scriptscriptstyle Q\bar Q{\rm J}}$ we denote the \FJ{} phase space, with $\Delta_{\rm pwg}$ the \POWHEG{} Sudakov form factor
featuring a default cutoff of $\LambdaPWG=0.89$\,GeV, and with $\Phi_{\tmop{rad}} $ and $\ptrad$ 
the phase space and the transverse momentum of the second radiation. $B_{\scriptscriptstyle Q\bar Q{\rm J}}$ and $R_{\scriptscriptstyle Q\bar Q{\rm J}}$ are the squared tree-level matrix elements for \FJ{} and \FJJ{} production, respectively.

In \eqn{eq:minnlo}, ${\rm d}\sigma^{(1,2)}_{\scriptscriptstyle Q\bar Q{\rm J}}$ denotes 
the differential cross section for \FJ{} production at  
first- and second-order, while the other terms stem from the 
analytical transverse-momentum resummation formula for $Q\bar Q$ production \cite{Zhu:2012ts,Li:2013mia,Catani:2014qha,Catani:2018mei,Ju:2022wia}, which 
can be written in the following symbolic form \cite{Mazzitelli:2020jio,Mazzitelli:2021mmm}:
\begin{align}
\label{eq:resum}
{\rm d}\sigma_{\scriptscriptstyle Q\bar Q}^{\rm res}&=\frac{{\rm d}}{{\rm d}\pt}\left\{\left[\sum_{i=1}^{n_c}\mathcal{C}_i\,e^{-S_i}\right]\mathcal{L}\right\}=\sum_{i=1}^{n_c}\mathcal{C}_i\,e^{-S_i}\underbrace{\left\{-S_i^\prime\,\mathcal{L}+\mathcal{L}^\prime\right\}}_{\equiv D_i}\,,
\end{align}
after some simplifications allowed within our desired accuracy (i.e.\ NNLO and preserving the accuracy of the parton shower).
This formula encodes all singular terms in the transverse momentum of 
the $Q\bar Q$ pair ($\pt$) up to relative $\as^2$ accuracy.
Note that the sum over partonic channels shall be understood as being implicit, while
the explicit sum over $n_c$ ($n_c=4$ for $q\bar{q}$ channels and $n_c=9$ for the $gg$ channel) originates from independent colour configurations.
This notation is required, since the 
logarithmic corrections arising from soft wide-angle exchanges between the final-state heavy quarks as well as final-initial state interferences
render the soft anomalous dimensions for heavy-quark pair production $\mathbf{\Gamma}^{(1)}_t$ to be matrix/operator in colour space. Thus,
through its exponentiation the Sudakov form factor, $e^{-S_i}$, becomes colour dependent as indicated by the subscript $i$.
The sum in $i$ is a consequence of diagonalizing $\mathbf{\Gamma}^{(1)}_t$, which generates the complex coefficients $\mathcal{C}_i$ 
that fulfil  $\sum_{i=1}^{n_c}\mathcal{C}_i=1$. 
We refer to \citere{Mazzitelli:2021mmm} for details. $\mathcal{L}$ is the 
luminosity factor including 
the squared hard-virtual matrix elements for $Q\bar Q$ production and 
the convolution of the collinear coefficient functions with the 
parton distribution functions (PDFs). $\mathcal{L}$
also receives additional contributions from azimuthal correlations of soft wide-angle exchanges 
between the heavy quarks and the initial state when taking the azimuthal average, 
which have been calculated in \citere{Catani:2023tby}.

The form of the $\bar{B}^{\,\rm MiNNLO_{\rm PS}}$ in \eqn{eq:minnlo}
follows directly from matching the singular terms described by the resummation 
formula in \eqn{eq:resum} with the fixed-order cross section 
${\rm d}\sigma^{(1,2)}_{\scriptscriptstyle Q\bar Q{\rm J}}$, while removing 
double counting and using a matching scheme where the Sudakov form factor 
is factored out:
\begin{align}
 & {\rm d}\sigma_{\scriptscriptstyle Q\bar Q}^{\rm res}+[{\rm d}\sigma_{\scriptscriptstyle Q\bar Q{\rm J}}]_{\rm f.o.}-[{\rm d}\sigma_{\scriptscriptstyle Q\bar Q}^{\rm res}]_{\rm f.o.}=\sum_{i=1}^{n_c}\mathcal{C}_i\,e^{-S_i}\bigg\{D_i+[{\rm d}\sigma_{\scriptscriptstyle Q\bar Q{\rm J}}]_{\rm f.o.}\,\underbrace{\frac{1}{[e^{-S_i}]_{\rm f.o.}\,}}_{1+S_i^{(1)}\cdots}\underbrace{\,\,-\,\frac{[{\rm d}\sigma_{\scriptscriptstyle Q\bar Q}^{\rm res}]_{\rm f.o.}\,\,}{[e^{-S_i}]_{\rm f.o.}}}_{-D_i^{(1)}-D_i^{(2)}\cdots}\bigg\}\\
&\quad\quad\quad\quad\quad\quad\quad\,\,\approx \sum_{i=1}^{n_c}\mathcal{C}_i\,e^{-S_i}\,\bigg\{{\rm d}\sigma^{(1)}_{\scriptscriptstyle Q\bar Q{\rm J}}\big(1+S_i^{(1)}\big)+{\rm d}\sigma^{(2)}_{\scriptscriptstyle Q\bar Q{\rm J}}+\underbrace{\left(D_i-D_i^{(1)}-D_i^{(2)}\right)}_{\equiv D_i^{(\ge 3)}}\bigg\}\,,\nonumber
\end{align}
where $[\cdots]_{\rm f.o.}$ yields the expansion up to a given fixed order in $\as$,
and $X^{(n)}$ is the $n$-th coefficient in the $\as$ expansions of $X$ including 
its coupling $\as^n$. 
The third-order term $D_i^{(\ge 3)}$ includes the relevant singular corrections
to reach NNLO QCD accuracy for inclusive \F{} production, which in \eqn{eq:minnlo}
are spread appropriately in the \FJ{} phase space through the factor $F^{\rm corr}$.
Regular contributions at this order are of subleading nature. 
Conversely, we define the \minlo{} cross section, which
correspond to a merging of $0$-jet and $1$-jet multiplicities at NLO QCD accuracy,
by excluding the NNLO $D_i^{(\ge 3)}$ corrections in \eqn{eq:minnlo}.\footnote{Note that the \minlo{} prediction for $b$-jet production are also a new result of the present paper.}

The \minnlo{} $b\bar{b}$ generator in the 4FS is implemented
 in the \POWHEGBOXRES{} framework~\cite{Jezo:2015aia}
 using the interface to \OpenLoops{}~\cite{Cascioli:2011va,Buccioni:2017yxi,Buccioni:2019sur}, 
which was developed in \citere{Jezo:2016ujg}, to obtain the tree-level and one-loop
amplitudes for $pp\to b\bar{b}$+jet process at NLO+PS in the 4FS with 
massive bottom quarks. Also the relevant $pp\to b\bar{b}$ amplitudes 
at tree-level and one-loop are evaluated through  \OpenLoops{}, while the 
two-loop amplitude is taken from the
numerical implementation in \citere{Barnreuther:2013qvf}, which were originally derived for top-quark pair production and appropriately extended to bottom-quark pair production in the 4FS.

Finally, we also include the contribution from massive four-bottom production, $pp\to b\bar b b\bar b$ , in our predictions. This contribution is formally of $\mathcal{O}(\alpha_s^2)$ relative to the Born process, but is separately finite and has been simulated using a dedicated LO+PS generator that we implemented within the \POWHEGBOXRES{} framework. This contribution is not included in the NLO \POWHEG{} simulation of $b\bar{b}$+jet, which includes only massless real radiation. Its inclusion can be relevant when comparing to $b$-jet measurements at the LHC, as it contributes to both inclusive and exclusive $b$-jet final states and therefore enters the experimental fiducial cross sections at the same perturbative order as the NNLO corrections.
Moreover, the $pp\to b\bar b b\bar b$ contribution also plays a role in the cancellation of $\log(m_b)$-enhanced terms. Such terms arise in virtual corrections and are, in principle, compensated by corresponding contributions from the real-emission $4b$ final state. However, the explicit inclusion of these terms demonstrates that an incomplete cancellation of these $\log(m_b)$ terms does not necessarily lead to sizeable numerical effects. Indeed, we find the $4b$ contribution to amount to less than $1\%$ of the NNLO+PS cross section for almost all observables. The only notable exception is the invariant-mass distribution of the two leading $b$-jets, where it reaches a few percent in the high-mass tail, while still remaining well below the residual perturbative scale uncertainty. We refer to the discussion around \fig{fig:atlas1} in \sct{sec:results}, where its impact is studied in more detail.

There is, however, one additional caveat concerning the $4b$
contribution. 
As discussed above, at the fixed-order, or equivalently at the Les-Houches-event (LHE) level, this contribution is found to have a small numerical effect.
However, when the $4b$
sample is showered as an independent LO+PS process, the default shower
starting scale is taken to be the partonic centre-of-mass energy. This
results in an unphysical enhancement of the high-energy tails of the
$b$-jet distributions by several orders of magnitude.
The effect occurs when the shower generates an excessive amount of radiation (effectively in ``power shower'' mode) which, during jet reconstruction, recombines with bottom quarks from the bulk of the events and migrates into the steeply
falling high-$p_T$ and high-$m_{bb}$ jet regions.
This behaviour is unphysical because, when viewed as part of the
NNLO+PS $b\bar b$ calculation, the $4b$ final state already contains
two additional hard emissions with respect to the underlying
$b\bar b$ process. Viewed in this way, subsequent shower emissions should be
generated at a substantially lower scale than the characteristic scale
of the $4b$ matrix element. 
Rather than using the partonic centre-of-mass energy as the starting scale, we instead select the minimum transverse momentum among all possible $b\bar b$ pairings in the $4b$ final state in a similar way to how the \POWHEG{} assigns the starting scale for real radiation in the massless case.
With this choice, the shower corrections remain at the level of $10$--$20\%$, yielding
physically reasonable predictions close to the original LHE-level
results, instead of the orders-of-magnitude enhancement obtained with
the default shower setup.

\section{Definition of jet flavour}\label{sec:flavour}

In experimental analyses performed at the LHC, there exist multiple approaches for the definition of jet flavour, see for example \citeres{LHCb:2015tna,ATLAS:2015thz,CMS:2017wtu,LHCb:2021dlw}.
The exact definition of how jet flavour is assigned varies for among the different experiments, and sometimes differs on an analysis-by-analysis basis within each experiment.
Of these variants, a commonly used approach is to define a jet as a $b$-jet if: there must be at least one reconstructed $B$-hadron within a small distance separation with respect to the jet axis, $\Delta R(B,{\rm jet}) < R_{\rm tag}$, where the usual definition of $\Delta R^2 = \Delta \phi_{ij}^2 + \Delta y_{ij}^2$ is assumed.
Often the $B$-hadron is also required satisfy a minimal transverse momentum requirement $p_{T,B} > p_{T,{\rm tag}}$.

Such a prescription cannot be directly applied to fixed-order calculations performed in the five-flavour scheme, where bottom quarks are treated as massless. 
In this case, the flavour assignment must be defined in an infrared- and collinear-safe (IRC-safe) manner to ensure the proper cancellation of singularities between real and virtual contributions.
In contrast, predictions obtained in a massive scheme, such as the 4FS, remain finite when the experimental jet-flavour definition is employed, since the bottom-quark mass regulates the collinear and soft singularities. 
However, the resulting cross sections contain logarithms of the form $\log(m_b)$, which arise from configurations that would become singular in the massless limit. Although these logarithms are formally finite, they can, in principle, become sizeable and deteriorate the perturbative convergence of the fixed-order expansion. Therefore, even for bottom-quark production in the 4FS, such as the $b\bar b$ production process considered in this work, it can be useful to employ jet-flavour definitions that avoid or reduce these potentially large logarithmic contributions.

Over the past few years, considerable effort has been devoted to developing jet-flavour definitions that remain IRC safe in massless schemes while preserving the desirable properties of the anti-$k_t$ jet algorithm. 
Several approaches have been proposed, including \textit{Flavoured anti-}$k_t$~\cite{Czakon:2022wam}, \textit{Flavour Dressing}~\cite{Gauld:2022lem}, and \textit{Interleaved Flavour Neutralisation} (IFN)~\cite{Caola:2023wpj}.

The origin of the problem is well understood. When a conventional flavour assignment is applied to fixed-order predictions in a massless scheme, certain QCD configurations alter the flavour assignment of a jet in a way that spoils the cancellation of infrared singularities between real and virtual contributions. 
Examples of configurations responsible for this behaviour are the collinear splitting of a gluon into a $b\bar b$ pair, with both quarks clustered into the same jet, and soft wide-angle emissions of bottom quarks that are subsequently clustered with an unrelated hard jet. 
As pointed out before, it is common in the experiments to require the transverse momentum of the $B$-hadron inside a $b$-jet to have a minimum transverse threshold, typically of the order of 5\,GeV. 
Such threshold also introduces a collinear sensitivity, since configurations that involve a hard collinear emission from the bottom quark will modify the momentum of the identified hadron.
A flavour definition that does not properly account for all of those configurations therefore leads to perturbatively ill-defined jet observables in a massless scheme, while in a massive scheme it gives rise to potentially large logarithmic contributions that may deteriorate the perturbative convergence.

In this work, our baseline is the 4FS \minnlo{} calculation interfaced with a fully exclusive parton shower, which facilitates the inclusion of hadronisation effects. This allows us to provide predictions for $b$-jet observables using the same $b$-jet flavour definitions as in the experimental analyses, and to compare them with predictions based on theoretically motivated $b$-jet definitions.

These results will be shown in the following section. Before doing so, we introduce the different versions of tagging procedures that we consider. 
In the following we assume the anti-$k_T$ algorithm as a baseline, and further assume stable $B$-hadrons as inputs to the jet reconstruction algorithm.
While we discuss tagging procedures with respect to $B$-hadrons, the corresponding procedures can also be applied at parton-level with bottom quarks.

\paragraph{Experimental tagging, {\tt exp}.} First, we consider an experimental-like tagging procedure in which a $b$-jet has flavour assigned when the reconstructed jet contains at least one $B$-hadron.
Notably, if the $b$-jet contains two or more $B$-hadrons, it is still considered as a $b$-jet.

We note that if $B$-hadrons were not considered as input particles to the jet reconstruction, but only their decay products,  
then an alternative criterion for the heavy flavour assignment (rather than the constituent one) is required.
For instance, this could be the distance measure of the reconstructed $B$-hadron and the jet axis, or a ghost matching criterion for the reconstructed $B$-hadron could be applied.

\paragraph{Experimental tagging with a $B$-hadron momentum threshold,  {\tt exp(thr)}.} As an adjustment of this {\tt exp} tagging procedure, one can additionally require that the $B$-hadron must also satisfy a minimal $p_T$ threshold, such as in the $b$-jet analyses by the ATLAS collaboration~\cite{ATLAS:2011ac,ATLAS:2016anw}. 
We refer to this tagging procedure as {\tt exp(thr)}.

\paragraph{Naive flavour tagging,  {\tt naive}.} 
As a simple modification of the {\tt exp} procedure, one can introduce a different flavour definition, in which one counts the number of bottom quarks (anti-quarks) $n_b$ ($n_{\bar b}$) of the jet constituents.
This approach assigns a non-zero jet flavour only when $n_b - n_{\bar b} \neq 0$. 
As it is experimentally challenging to distinguish bottom quarks and anti-bottom flavoured hadrons, 
we apply a modification of this approach where no distinction between $n_b$ and $n_{\bar b}$ is made and a non-zero jet flavour assignment is applied when a jet contains an odd number of $b$-flavoured objects.
In the following, we refer to that approach as ``{\tt naive}" tagging.

These prescriptions remove the flavour ambiguity associated with collinear $g\to b\bar b$ splittings, since a collinear bottom-quark pair carries zero net flavour. Although they do not cure the IRC sensitivity arising from soft wide-angle heavy-quark emissions, they eliminate the dominant and phenomenologically most relevant source of flavour misidentification.

\paragraph{Interleaved Flavour Neutralisation, IFN.} As an example of an IRC-safe definition of jet flavour, we consider the \textit{Interleaved Flavour Neutralisation} (IFN)~\cite{Caola:2023wpj} algorithm, labelled \texttt{IFN}.
In short, this algorithm incorporates flavour information directly into the clustering sequence through a neutralisation procedure that systematically removes the problematic soft and collinear flavour configurations. Throughout this work we employ the default parameter choices recommended in Ref.~\cite{Caola:2023wpj}, namely $\alpha=2$ and $\omega=1$ for the neutralisation distance.
Jets with an even number of $b$-flavoured objects are not considered $b$-jets, as in the {\tt naive} tagging
defined above.

\paragraph{Flavour dressing.} As an alternative approach of an IRC-safe jet-flavour assignment, we consider the \texttt{Flavour dressing} approach~\cite{Gauld:2022lem}. This approaches differs to IFN in that it requires that a set of flavour-blind jets are reconstructed first (such as anti-$k_T$ jets), and their flavour is then assigned in a later stage. The flavour assignment makes use of an association criterion between flavoured particles and the jets, as well as a modified version of the flavour-$k_T$ distance measure as introduced in \citere{Caola:2023wpj}. Again, jets with an even number of $b$-flavoured objects are not considered $b$-jets, as in the {\tt naive} and {\tt IFN} taggings
defined above.

\paragraph{Additional comments:} 
Firstly, we re-iterate that we consider stable $B$-hadrons in our simulations, and use these particles as inputs to the jet reconstruction.
Secondly, with the exception of the {\tt exp(thr)} approach, it is assumed that $B$-hadrons can be reconstructed for all $p_{T,B}$ values (i.e.\ down to vanishing $p_{T,B}$).
This is clearly challenging in an experimental context, and means that some level of unfolding (or an equivalent procedure) is required to apply such a definition in measurement. 
This introduces some level of dependence on the description of the small-$p_{T,B}$ region in the measurement through the simulation (e.g.\ the available Monte Carlo Parton Shower tools).
Furthermore, the {\tt naive} and the considered IRC-safe algorithms require to distinguish between jets that contain one or more $B$-hadrons, which is very challenging in an experimental context.
Similarly, other criteria required to render $b$-jets IRC safe are not straightforward to achieve experimentally either.
Continued support for this experimental effort~\cite{ATLAS:2018zhf} is crucial so that state-of-the-art theoretical predictions (specifically, at fixed order in a massless scheme) can be used, and uncertainties arising from unfolding can be reduced.

\section{Phenomenological results}\label{sec:results}

In this section, we present phenomenological results for $b$-jet production at the LHC at different center-of-mass energies.
In particular, we study the relevance of using different flavour-aware jet-clustering algorithms, and we 
compare our NNLO+PS predictions against four different experimental measurements: two 7\,TeV measurements by ATLAS \cite{ATLAS:2011ac,ATLAS:2016anw}, a 7\,TeV measurement by CMS \cite{CMS:2012pgw}
and a 13\,TeV measurement by LHCb \cite{LHCb:2020frr}. We refer to those publications for the definition of the respective fiducial phase spaces and definition of the $b$-jets.

Since we employ the 4FS throughout our calculation, bottom quarks are treated as being massive and we set their 
pole mass to $m_b=4.92$\,GeV \cite{LHCHiggsCrossSectionWorkingGroup:2016ypw}. For the PDFs we choose the NNLO set of 
NNPDF3.1~\cite{Ball:2017nwa} consistent with $N_f=4$ number of light quark flavours (specifically {\tt NNPDF31\char`_nnlo\char`_as\char`_0118\char`_nf\char`_4})
and the strong coupling with 4FS running corresponding to that set.
We use the {\sc lhapdf} interface \cite{Buckley:2014ana} to read the PDF set and pass them to the {\sc hoppet} code \cite{Salam:2008qg} to evolve them to different scales and 
to perform relevant convolutions. The renormalization~($\muR$) setting of the strong coupling constants  
relative to the Born level and factorization ($\muF$) scale setting are intrinsic to the \minlo{}/\minnlo{} method, which has been 
described in detail in section 4.3 of \citere{Mazzitelli:2021mmm}.
The scales for the two overall powers of $\as$ at Born level, on the other hand, can be set freely and we choose them as
\begin{align}
\muRc=\KR\,\frac{H_T}{2},\quad\text{with}\quad 
H_T = \sum_{i=1}^{n_B} m_{T,i} \,,
\end{align}
where the sum runs over all $n_B$ (underlying) Born particles and and $m_{T,i}$ is the transverse mass  of particle $i$, namely  $m_{T,i} = \sqrt{p_{T,i}^2+m_i^2}$.
To estimate the uncertainties related to missing higher-order contributions, we use $7$-point scale variations, i.e.\ varying $\KR$ and $\KF$ for all scales by a factor of two in each direction with the constraint $1/2\le \KR/\KF\le 2$.
As far as the modified logarithms are concerned, we follow the definition given in \citere{Mazzitelli:2021mmm}, 
which consistently switches off all resummation effects at large transverse momenta, where
we use the standard scale choice of $Q=m_{b\bar b}/2$. Also the other technical settings are taken as in \citere{Mazzitelli:2021mmm} ($Q_0 = 2$\,GeV, $\KR = \KF = 1$ for the central scales). 

The \PYTHIA{8} \cite{Sjostrand:2014zea} parton shower is used throughout, with the Monash 2013 tune \cite{Skands:2014pea},
where (unless stated otherwise) we have turned on effects from hadronization and from multi-parton interactions (MPI) 
to obtain a fully realistic simulation of bottom-flavoured jets.
We note that all current hadronization tunes were carried out using calculations
with a perturbative accuracy lower than NNLO. In the future, it would be 
interesting to obtain new tunes using NNLO Monte Carlo generators.
\begin{figure}[t!]
\begin{center}
\begin{tabular}{cc}
\includegraphics[width=.42\textwidth, page=1]{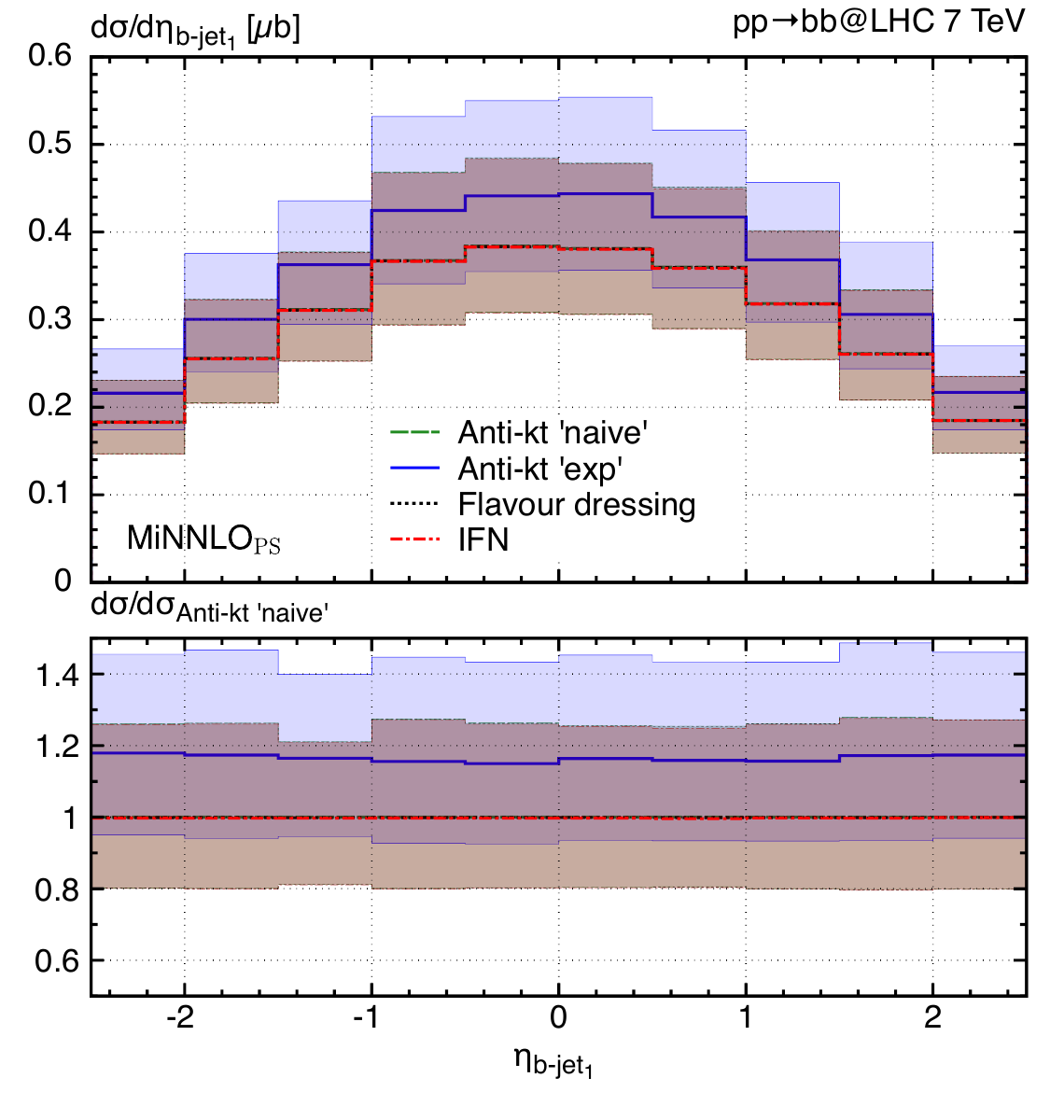}
&
\hspace{0.25cm}
\includegraphics[width=.42\textwidth, page=3]{plots/all_plots_plus4b.pdf}\\
\includegraphics[width=.42\textwidth, page=7]{plots/all_plots_plus4b.pdf}
&
\hspace{0.25cm}
\includegraphics[width=.42\textwidth, page=9]{plots/all_plots_plus4b.pdf}
\end{tabular}
\vspace*{1ex}
\caption{\label{fig:jetalgo} Comparison of different jet-flavour definitions obtained with our \minnlo{} 4FS generator, including the contribution from massive four-bottom production. See text for more details.}
\end{center}
\end{figure}

As will be evident in the following phenomenological studies (for both 7 and 13~TeV hadronic centre-of-mass energies), the differential cross-section predictions for both one-$b$-jet or two-$b$-jet-based measurements cover many orders in magnitude and often finely binned predictions are required to match the experimental measurements.
Since bottom-quark pair production is dominated by the kinematic region which is not far from the $b\bar b$ threshold, typically no or very few events are generated in the required kinematic regions for the $b$-jet analyses.
We overcome this issue by introducing a combination of Born suppression factors to increase the sampling of the high-transverse-momentum region, and, where suitable, generation cuts.
Despite this, it was still numerically challenging to populate events in some of the more extreme kinematic regions.
As a related point, for comparison with the \minnlo{} prediction we have also generated dijet events at NLO+PS accuracy in the 5FS. 
The generation of events in that case was also computationally expensive: While the actual CPU cost per event is far below that of the massive \minnlo{} simulation, the component of the full dijet 5FS cross section that leads to $b$-jets is only a small fraction ($\approx 3\%$, depending on the kinematic regime).

We now start by comparing the different $b$-jet definitions introduced in the previous section, obtained with \minnlo{}. Figure~\ref{fig:jetalgo} shows the pseudorapidity ($\eta_{b\text{-jet}_1}$), and transverse momentum  ($p_{T,b\text{-jet}_1}$) of the hardest $b$-jet, together with the invariant mass ($m_{b\bar b}$) and transverse momentum ($p_{T,b\bar b}$) of the two hardest $b$-jets.
For all algorithms we use $R=0.4$ and require a $b$-jet to have at least transverse momentum of $p_{T,b\text{-jet}}\ge 30$\,GeV and a rapidity of $|\eta_{b\text{-jet}}|\le 2.5$, while no other cuts are applied (in particular, no threshold on the $B$-hadron momenta is applied, c.f.\ \sct{sec:flavour}).
We observe that the dominant differences between the various flavour definitions arise from the treatment of collinear $g\to b\bar b$ splittings. This is evident from the comparison of the {\tt exp} and {\tt naive} definitions, where the latter differs only through the removal of contributions from collinear $g\to b\bar b$ splittings through the modulo-two flavour assignment. The resulting effect, of about 20\%, is essentially independent of the jet pseudorapidity, while it increases steadily in the high-energy tails of the remaining observables. In particular, for the transverse momentum of the hardest $b$-jet, the {\tt exp} definition predicts cross sections that are up to a factor of two larger than those obtained with the other flavour definitions at large $p_T$ ($p_{T,b\text{-jet}_1}\gtrsim 200 $GeV). Similar, albeit somewhat smaller, effects are observed for the dijet invariant mass and the transverse momentum of the $b\bar b$ system.
Comparing the IRC-safe flavour definitions {\tt Flavour dressing} and {\tt IFN} with the {\tt naive} prescription, we find only very small differences. This demonstrates that the dominant numerical effect is already captured by removing collinear $g\to b\bar b$ splittings, while the soft wide-angle configurations responsible for the remaining IRC unsafety have only a minor impact. Noticeable deviations are observed only in the far tails of the $p_{T,b\text{-jet}_1}$ and $p_{T,b\bar b}$ distributions, where they reach about ten percent. These effects remain well below the residual perturbative scale uncertainties and are therefore currently phenomenologically negligible.

\begin{table}[t]
  \centering
  \renewcommand{\arraystretch}{1.25}
  \begin{tabular}{lccc}
    \hline
    Observable & Data & \minlo{} & \minnlo{} \\
    \hline

    \multicolumn{4}{c}{\textbf{CMS} $\sqrt{s}=7~\mathrm{TeV}$~
      \cite{CMS:2012pgw}} \\

    $\sigma(b\text{-jet},\,p_{T,b\text{-jet}}>18~\mathrm{GeV})$
      & $9.75\,
        {\scriptstyle
          \pm 0.32_{(\mathrm{stat})}
          \pm 1.67_{(\mathrm{syst})}
          \pm 0.39_{(\mathrm{lum})}}
        ~\mu\mathrm{b}$
      & $9.37(10)\,{}^{+35.8\%}_{-24.3\%}$
      & $10.19(9)\,{}^{+25.7\%}_{-19.0\%}$ \\

    $\sigma(b\text{-jet},\,p_{T,b\text{-jet}}>32~\mathrm{GeV})$
      & $1.73\,
        {\scriptstyle
          \pm 0.07_{(\mathrm{stat})}
          \pm 0.20_{(\mathrm{syst})}
          \pm 0.07_{(\mathrm{lum})}}
        ~\mu\mathrm{b}$
      & $1.429(8)\,{}^{+31.6\%}_{-23.0\%}$
      & $1.563(8)\,{}^{+24.1\%}_{-19.1\%}$ \\
    \hline

    \multicolumn{4}{c}{\textbf{LHCb} $\sqrt{s}=13~\mathrm{TeV}$~
      \cite{LHCb:2020frr}} \\

    $\sigma(pp\to b\bar b\text{-dijet}+X)$
      & $53.0\,
        {\scriptstyle
          \pm 9.5_{(\mathrm{stat+syst})}
          \pm 2.1_{(\mathrm{lum})}}
        ~\mathrm{nb}$
      & $89.9(1.2)\,{}^{+31.9\%}_{-21.6\%}$
      & $112.7(1.2)\,{}^{+11.2\%}_{-9.3\%}$ \\
    \hline
  \end{tabular}
  \vspace{0.35cm}
  \caption{Comparison of measured bottom-jet cross sections with \minlo{} and
  \minnlo{} predictions. For the CMS measurements, the data uncertainties are
  statistical, systematic, and luminosity uncertainties, respectively. For the
  LHCb measurement, the first uncertainty combines the statistical and
  systematic components, while the second is due to the luminosity. Theory
  uncertainties are obtained from scale variations, and Monte Carlo integration
  uncertainties are given in parentheses. See text for more details.}
  \label{tab:bjet-cross-sections}
\end{table}

We continue by comparing our predictions for $b$-jet production with LHC measurements. Throughout this section, we employ the standard anti-$k_T$ definition of $b$-jets, which has been used in all experimental analyses to date. In \tab{tab:bjet-cross-sections} we compare integrated fiducial $b$-jet cross sections measured by CMS \cite{CMS:2012pgw} and LHCb \cite{LHCb:2020frr} with \minlo{} and \minnlo{} predictions.\footnote{The relevant ATLAS analyses \cite{ATLAS:2011ac,ATLAS:2016anw} do not provide integrated $b$-jet cross sections.} The CMS analysis measures the inclusive $b$-jet cross section for two transverse-momentum thresholds, $p_{T,b\text{-jet}}>18$\,GeV and $p_{T,b\text{-jet}}>32$\,GeV, where the measured cross section is the sum over selected $b$-jets, such that events with multiple selected $b$-jets contribute multiple times. By contrast, the LHCb measurement corresponds to the cross section for events containing at least two $b$-jets.

For the CMS measurements, \minnlo{} provides an excellent description of the inclusive $b$-jet cross sections, with the central predictions lying within the experimental uncertainties. The corresponding \minlo{} predictions are systematically lower, but remain compatible with the data once experimental and theoretical uncertainties are taken into account. This trend is consistent with the comparison to the 4FS MC@NLO predictions presented in \citere{CMS:2012pgw}, which yield even smaller cross sections together with larger scale uncertainties. These comparisons demonstrate the importance of including NNLO corrections through \minnlo{} for precision predictions of $b$-jet production at the LHC. As expected, \minnlo{} also consistently reduces the perturbative scale uncertainties with respect to \minlo{} (and, more generally, 4FS NLO+PS) for all observables considered.

\begin{figure}[p]
\begin{center}
\begin{tabular}{cc}\vspace*{-0.3cm}
\includegraphics[width=.42\textwidth, page=1]{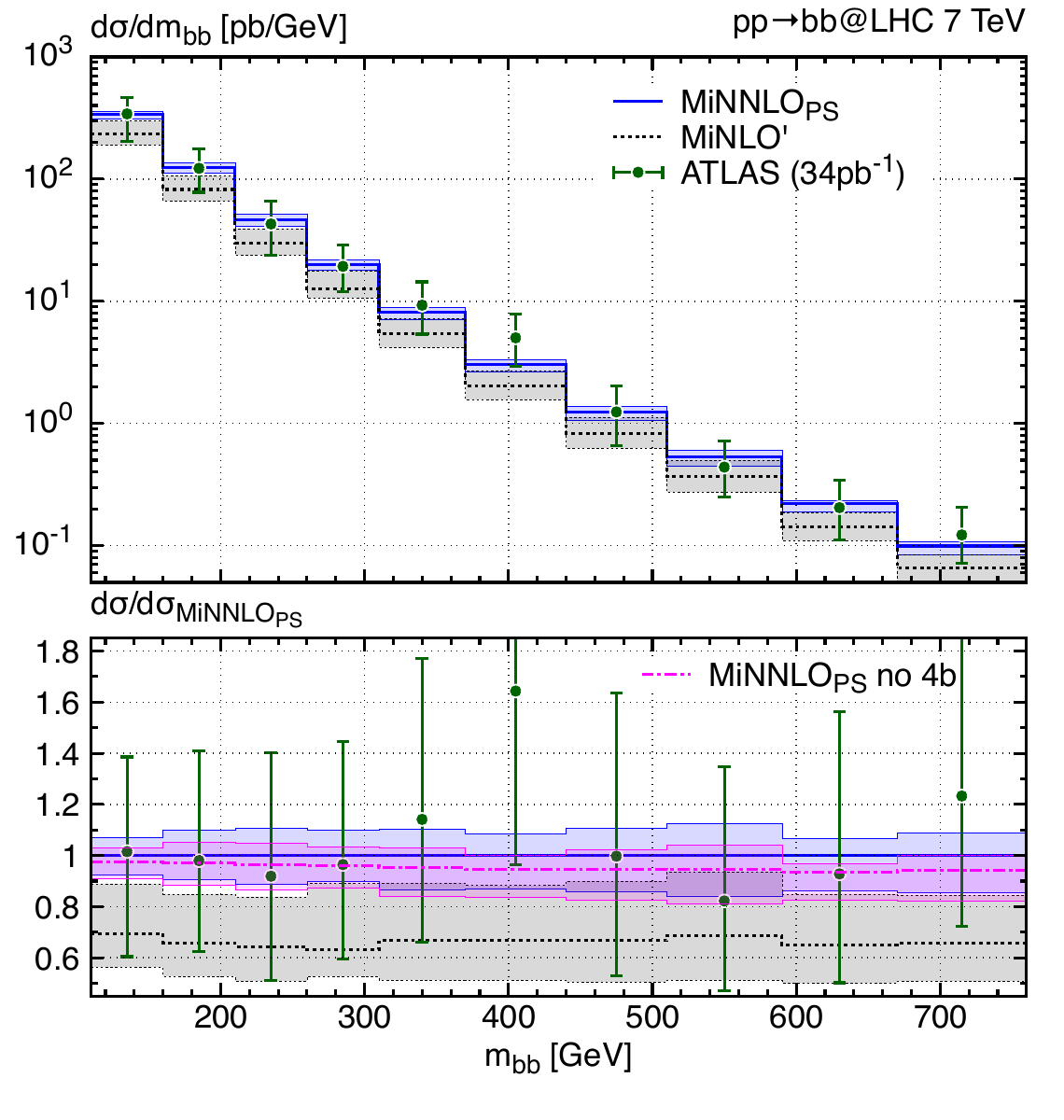}
&
\includegraphics[width=.42\textwidth, page=5]{plots/ATLAS_bjets.pdf}\\\vspace*{-0.3cm}
\includegraphics[width=.42\textwidth, page=4]{plots/ATLAS_bjets.pdf}
&
\includegraphics[width=.42\textwidth, page=2]{plots/ATLAS_bjets.pdf}\\
\includegraphics[width=.42\textwidth, page=3]{plots/ATLAS_bjets.pdf}
&
\includegraphics[width=.42\textwidth, page=6]{plots/ATLAS_bjets.pdf}
\end{tabular}
\vspace*{1ex}
\caption{\label{fig:atlas1} Comparison to 7\,TeV ATLAS data from Ref.~\cite{ATLAS:2011ac}. See text for details.}
\end{center}
\end{figure}

\begin{figure}[t!]
\begin{center}
\begin{tabular}{cc}
\includegraphics[width=.42\textwidth, page=3]{plots/LHCb_bjets.pdf}
&
\hspace{0.25cm}
\includegraphics[width=.42\textwidth, page=1]{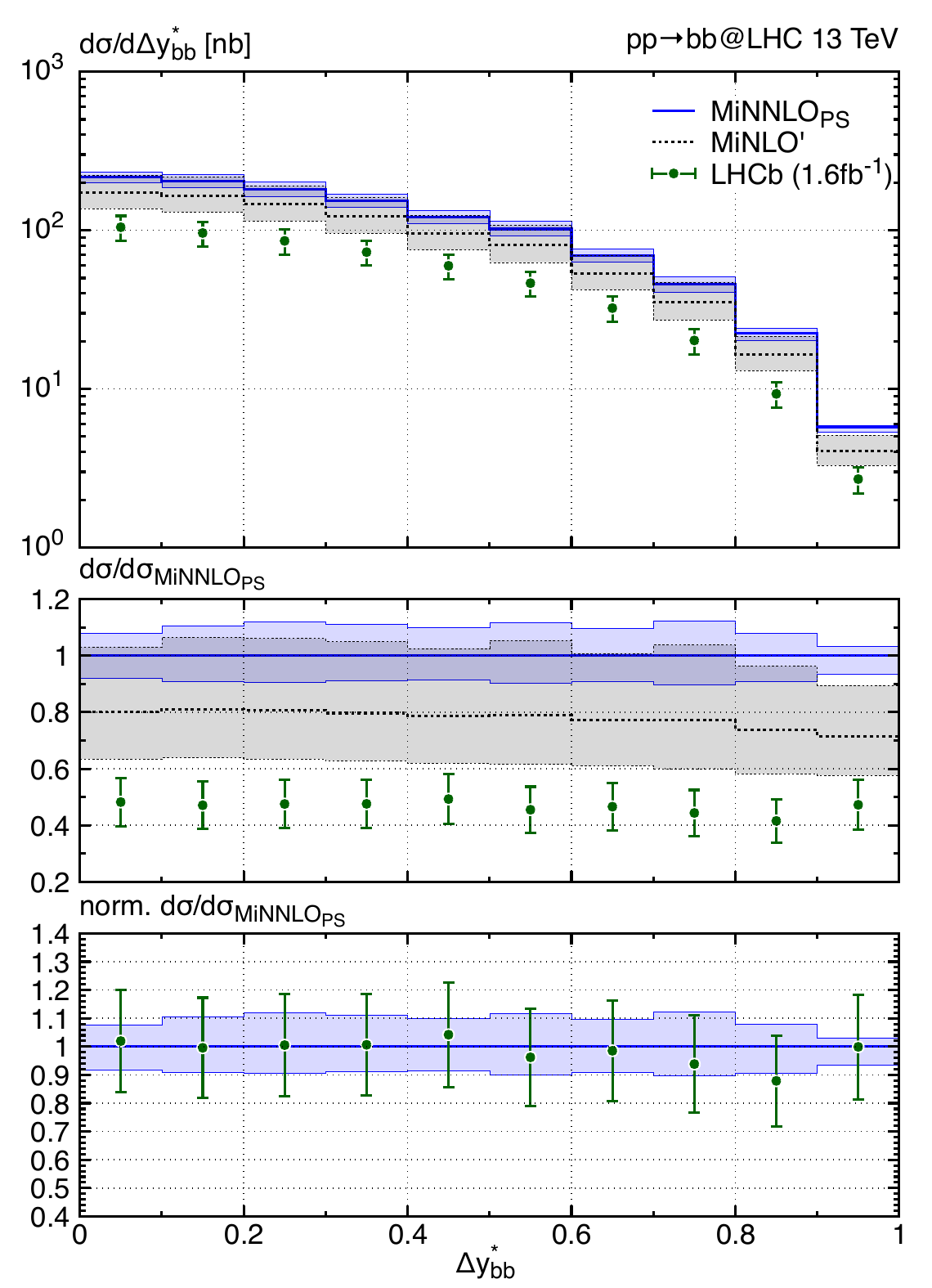}\\
\includegraphics[width=.42\textwidth, page=4]{plots/LHCb_bjets.pdf}
&
\hspace{0.25cm}
\includegraphics[width=.42\textwidth, page=2]{plots/LHCb_bjets.pdf}\\
\end{tabular}
\vspace*{1ex}
\caption{\label{fig:lhcb} Comparison to 13\,TeV LHCb data of Ref.~\cite{LHCb:2020frr}. See text for details.}
\end{center}
\end{figure}

The comparison with the LHCb measurement of the two-$b$-jet cross section paints a different picture. Both \minlo{} and \minnlo{} overestimate the measured cross section, with the latter exceeding the central experimental value by almost exactly a factor of two. We note that the 4FS MG5\_aMC@NLO prediction used for comparison in \citere{LHCb:2020frr} also overshoots the data, while \minnlo{}, as expected, adds sizeable positive NNLO corrections on top of the NLO+PS result. Despite extensive cross-checks of both the analysis implementation and the theoretical predictions, we have not been able to identify the origin of this discrepancy. 
In the absence of a definitive explanation, we simply report this discrepancy. For the differential distributions shown in \fig{fig:lhcb}, we therefore present both the absolute predictions, where the factor-of-two mismatch remains visible, and normalized distributions, which allow for a more meaningful comparison of the shapes.

We now turn to the comparison of differential predictions with LHC measurements. In the following figures, the main panels show the \minlo{} (black, dotted histograms) and \minnlo{} (blue, solid histograms) predictions together with the corresponding experimental measurement (green data points). We begin with the ATLAS measurement at 7\,TeV \cite{ATLAS:2016anw}, shown in \fig{fig:atlas1}, which displays the invariant-mass distribution of the two leading $b$-jets ($m_{bb}$) and the inclusive transverse-momentum spectrum of the $b$-jets ($p_{T,b\text{-jet}}$), the latter observable is shown both integrated over the full rapidity range ($|y_{b\text{-jet}}|\le2.1$) and in separate rapidity bins. 
We note that the $m_{bb}$ distribution is the standard two-$b$-jet
observable, whereas the $p_{T,b\text{-jet}}$ distributions are inclusive
in the number of selected $b$-jets, i.e.\ all $b$-jets satisfying the
respective rapidity selection contribute to the spectrum. To illustrate
the impact of the $4b$ contribution, the ratio panel of the $m_{bb}$
distribution also compares our default \minnlo{} prediction with the
corresponding result obtained without the $4b$ contribution (magenta,
dash-dotted histograms). The $4b$ contribution increases the cross
section by up to approximately $5\%$ in the high-$m_{bb}$ tail. For all
other distributions, its effect is too small to be resolved in the
plots, and we therefore omit the corresponding prediction in the
following figures.
For the $m_{bb}$ distribution, \minnlo{} yields sizeable NNLO QCD
corrections of about $30\%$ with respect to \minlo{}, lying near the
upper edge of the \minlo{} scale-uncertainty band, while reducing the
perturbative scale uncertainties by up to a factor of two, depending on
the invariant-mass region.
The \minnlo{} prediction provides an excellent description of the measured $m_{bb}$ spectrum, with deviations well below the experimental uncertainties, and represents a clear improvement over \minlo{}. By contrast, the inclusive $p_{T,b\text{-jet}}$ spectra receive more moderate NNLO QCD corrections of roughly $10\%$, while the scale-uncertainty bands of the \minlo{} and \minnlo{} predictions remain of comparable size. Also in this case, the \minnlo{} predictions are in very good agreement with the data, with nearly all measurements lying within one standard deviation and all within two standard deviations of the theoretical prediction.

\begin{figure}[p]
\begin{center}
\begin{tabular}{cc}\vspace*{-0.3cm}
\includegraphics[width=.42\textwidth, page=1]{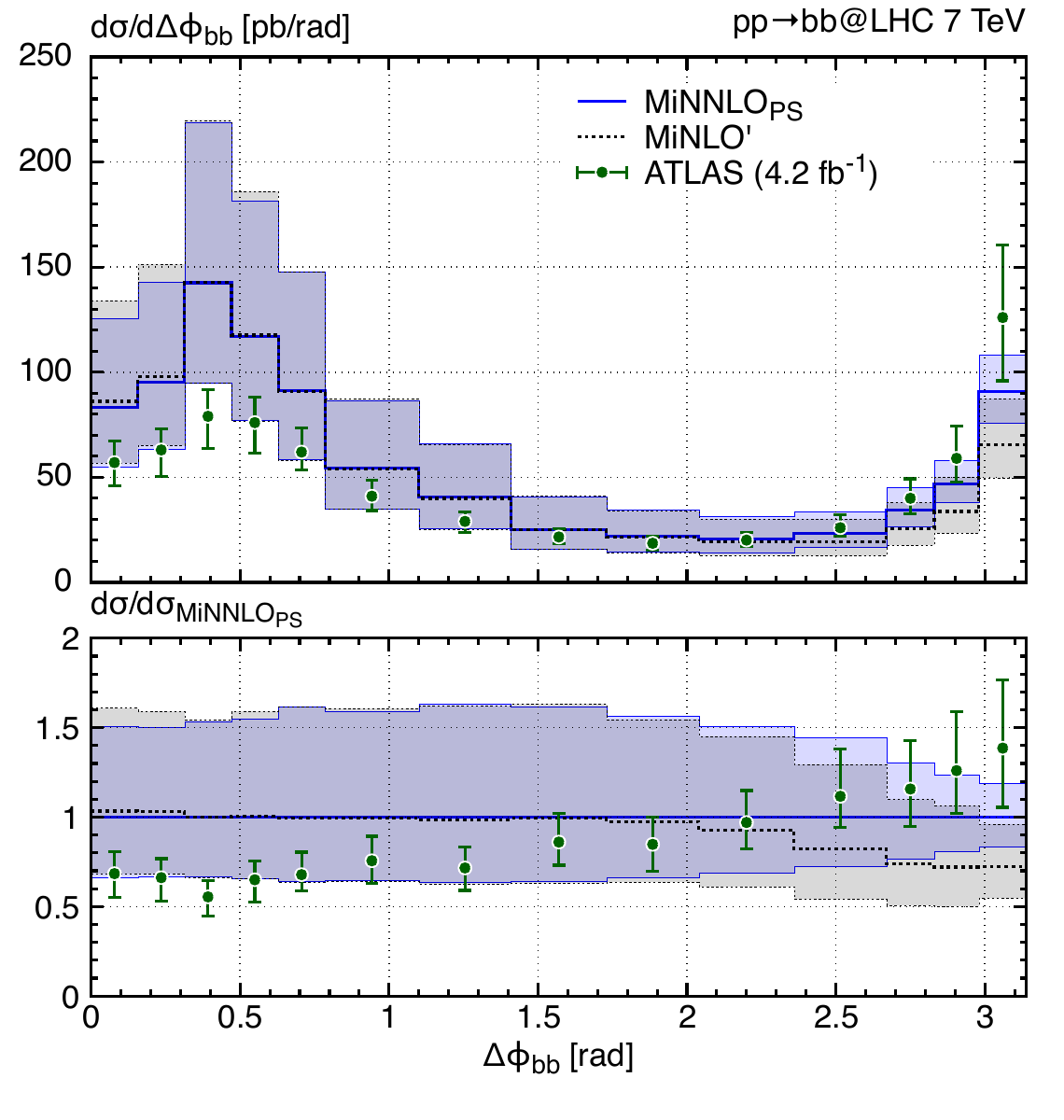}
&
\includegraphics[width=.42\textwidth, page=2]{plots/ATLAS_2_bjets.pdf}\\\vspace*{-0.3cm}
\includegraphics[width=.42\textwidth, page=3]{plots/ATLAS_2_bjets.pdf}
&
\includegraphics[width=.42\textwidth, page=4]{plots/ATLAS_2_bjets.pdf}\\
\includegraphics[width=.42\textwidth, page=5]{plots/ATLAS_2_bjets.pdf}
&
\includegraphics[width=.42\textwidth, page=6]{plots/ATLAS_2_bjets.pdf}
\end{tabular}
\vspace*{1ex}
\caption{\label{fig:atlas2} Comparison to 7\,TeV ATLAS data of Ref.~\cite{ATLAS:2016anw}. See text for details.}
\end{center}
\end{figure}

Next, we consider the more recent ATLAS 7\,TeV analysis of \citere{ATLAS:2016anw}, which measures a broader set of two-$b$-jet observables using a substantially larger integrated luminosity. The corresponding comparison is shown in \fig{fig:atlas2}, which displays the distributions of the azimuthal separation of the two leading $b$-jets ($\Delta\phi_{bb}$), their angular distance in the $\eta$–$\phi$ plane ($\Delta R_{bb}$), their invariant mass ($m_{bb}$), their transverse momentum ($p_{T,bb}$), as well as the rapidity sum $y_B=\frac12|y_{b\text{-jet}_1}+y_{b\text{-jet}2}|$ and rapidity difference $\Delta y_{bb}^*=\frac12|y_{b\text{-jet}_1}-y_{b\text{-jet}2}|$. This analysis requires the leading jet to satisfy $p_{T,\text{jet}_1}>270$\,GeV, thereby significantly restricting the available phase space.

Consequently, the formal perturbative accuracy is reduced in most regions of phase space, and \minnlo{} provides an accuracy beyond \minlo{} only in the region where the leading $b$-jet recoils against the second $b$-jet. 
This expectation is reflected in the scale-uncertainty bands, which remain of comparable size, amounting to roughly $\pm\mathcal{O}(30\text{–}50\%)$, and in the close agreement between the \minlo{} and \minnlo{} predictions over most of the phase space. Noticeable differences are observed only at large invariant masses and large separations of the two leading $b$-jets, where \minnlo{} yields larger corrections and reduced scale uncertainties, as expected.
Overall, the \minnlo{} predictions are in good agreement with the ATLAS data. Most data points are described within one standard deviation, while only a few show deviations at the level of two to three standard deviations. The shapes of the $m_{bb}$ and $p_{T,bb}$ distributions are, however, less well reproduced than those of the other observables. The theoretical uncertainties are sizable and exceed the experimental uncertainties. This can be attributed to the relatively high $p_{T,\text{jet}_1}$ cut, which effectively reduces the perturbative accuracy of the $p_{T,bb}$ distribution at intermediate and large values, and $m_{bb}$ distribution at intermediate and small values by one perturbative order.

\begin{figure}[t!]
\begin{center}
\hspace*{-0.5cm}
\begin{tabular}{ccc}
\includegraphics[width=.355\textwidth, page=2]{plots/CMS_bjets.pdf}
&
\hspace{-0.8cm}
\includegraphics[width=.355\textwidth, page=1]{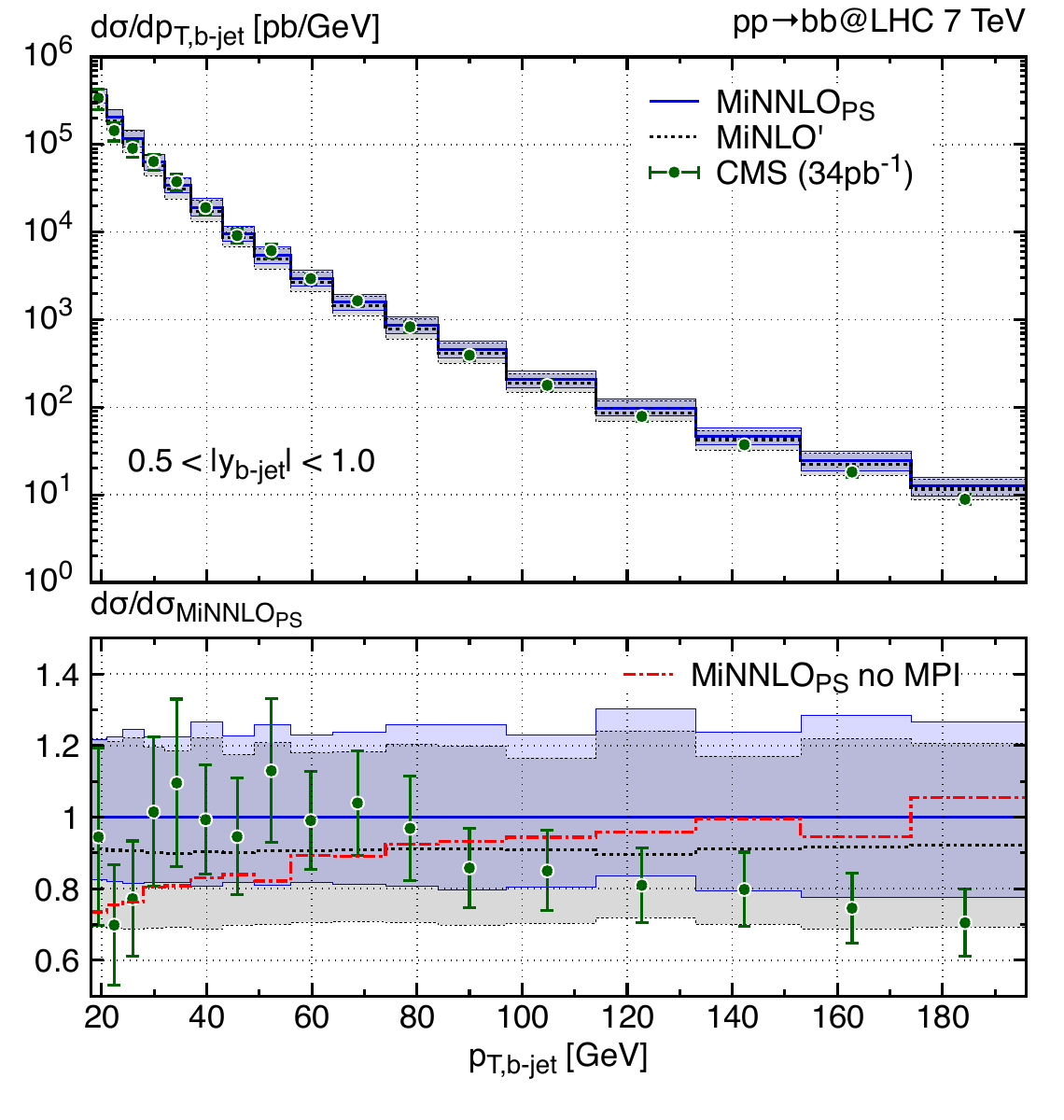}
&
\hspace{-0.8cm}
\includegraphics[width=.355\textwidth, page=4]{plots/CMS_bjets.pdf}\\
\includegraphics[width=.355\textwidth, page=3]{plots/CMS_bjets.pdf}
&
\hspace{-0.8cm}
\includegraphics[width=.355\textwidth, page=5]{plots/CMS_bjets.pdf}
\end{tabular}
\vspace*{1ex}
\caption{\label{fig:cms} Comparison to 7\,TeV CMS data of Ref.~\cite{CMS:2012pgw}. See text for details.}
\end{center}
\end{figure}

Also CMS has presented a differential measurement of the inclusive $p_{T,b\text{-jet}}$ spectrum in different bins of $y_{b\text{-jet}}$ at 
$7$\,TeV \cite{CMS:2012pgw}.
The corresponding comparison with our predictions is shown in \fig{fig:cms}. In the ratio panels, we additionally show the effect of MPI by including the \minnlo{} prediction obtained without MPI.
The MPI contribution reaches approximately $+30\%$ in the lowest $p_{T,b\text{-jet}}$ bins and decreases smoothly towards larger $p_{T,b\text{-jet}}$. 
The increased QCD activity introduced through the inclusion of MPI leads to a positive shift in the jet $p_{T}$.
This has more pronounced effect at relatively smaller $p_{T}$ values where the distribution falls more steeply.
This underscores the importance of employing full-fledged parton-shower Monte Carlo simulations for reliable predictions of $b$-jet observables.
As for the ATLAS inclusive $p_{T,b\text{-jet}}$ measurement discussed above, \minnlo{} yields relatively uniform NNLO QCD corrections of approximately $+10\%$ with respect to \minlo{}, while the scale uncertainties remain comparable in size.
Overall, the agreement with the CMS data is very good, with the majority of the measurements lying within one standard deviation of the \minnlo{} prediction and nearly all remaining points within two standard deviations. We note, however, that the relative theoretical uncertainties are larger than the experimental errors in the tails of the $p_{T,b\text{-jet}}$ distributions.

Finally, we consider the recent 13\,TeV measurement presented by LHCb in \citere{LHCb:2020frr} of differential two-$b$-jet production. 
The corresponding comparison to our theoretical predictions is shown in \fig{fig:lhcb}, which displays the $m_{bb}$ and $\Delta y^{*}_{bb}$ distributions, and the transverse-momentum spectrum of the leading $b$-jet ($p_{T,b\text{-jet}_1}$) and its pseudo-rapidity distribution ($\eta_{b\text{-jet}_1}$). 
In each plot, we include a second ratio panel showing the ratio of the normalized data and \minnlo{} distributions. This facilitates a comparison of their shapes, since, as already noted in the discussion of \tab{tab:bjet-cross-sections}, the absolute normalization differs by approximately a factor of two.
We observe that \minnlo{} induces sizable corrections of approximately $+20$--$30\%$ with respect to \minlo{}, accompanied by a significant reduction of the perturbative scale uncertainties. In the comparison with the absolute distributions, the factor-of-two offset between the data and the \minnlo{} predictions is clearly visible, consistent with the result reported in \tab{tab:bjet-cross-sections}. For the normalized distributions shown in the second ratio panels, however, the agreement with \minnlo{} is excellent: all data points are described within the experimental uncertainties, with the sole exception of the last bin of the $m_{bb}$ distribution. The remaining observables are reproduced very well, indicating that \minnlo{} provides an excellent description of the measured shapes.

Given that we do not find agreement with the LHCb data for the absolute value of the various distributions (with theory being roughly a factor of two larger, as discussed above), we have also performed a fixed-order study of the LHCb fiducial cross sections at NLO QCD accuracy, as an independent cross-check of our results.\footnote{As a side note, the NLO electroweak corrections to $b$-jet production at LHCb (in the 4FS) have been considered in \citere{Gauld:2019doc} are typically percent level, and hence can not account for the large differences.}
In doing so, we observed a peculiarity related to the choice of jet selection cuts used by LHCb to define the fiducial  region of the measurement.
The fiducial volume of the measurement is defined by requiring the presence of two R=0.5 anti-$k_T$ jets which have been tagged according to the {\tt exp} criterion.
These jets must further satisfy the kinematic constraints:
\begin{align}
p_{T,j} > 20{\rm~GeV}\,,\quad 2.2 < \eta_{j} < 4.2\,.
\end{align}
Importantly, `symmetric' cuts are applied to both the leading and sub-leading jet that form the dijet $bb$ system.

It has been pointed out in the past~\cite{Klasen:1995xe,Harris:1997hz,Frixione:1997ks} that such symmetric selections can lead to an IR sensitivity of the defined cross section, meaning that fixed-order predictions cannot be reliably applied. 
This issue is related to the exact back-to-back nature of the required final-state objects (in this case the $b$-jets).

To quantify whether this effect is relevant for the LHCb setup,
we consider again the set of jet selections as required by LHCb~\cite{LHCb:2020frr} (which we refer to as symmetric cuts) and study the invariant mass distribution of the reconstructed two $b$-jet system.
In addition, we also introduce a set of `asymmetric cuts' on the jets, namely the leading and subleading jets must require
\begin{align}
p_{T,j_1} > 20{\rm~GeV}\,,\quad
p_{T,j_2} > 17{\rm~GeV}\,,\quad 
2.2 < \eta_{j} < 4.2\,.
\end{align}
The difference with respect to the symmetric cuts is that we allow the sub-leading $b$-jet to have a slightly lower threshold.

\begin{figure}[t!]
\begin{center}
\begin{tabular}{cc}
\includegraphics[width=.42\textwidth]
{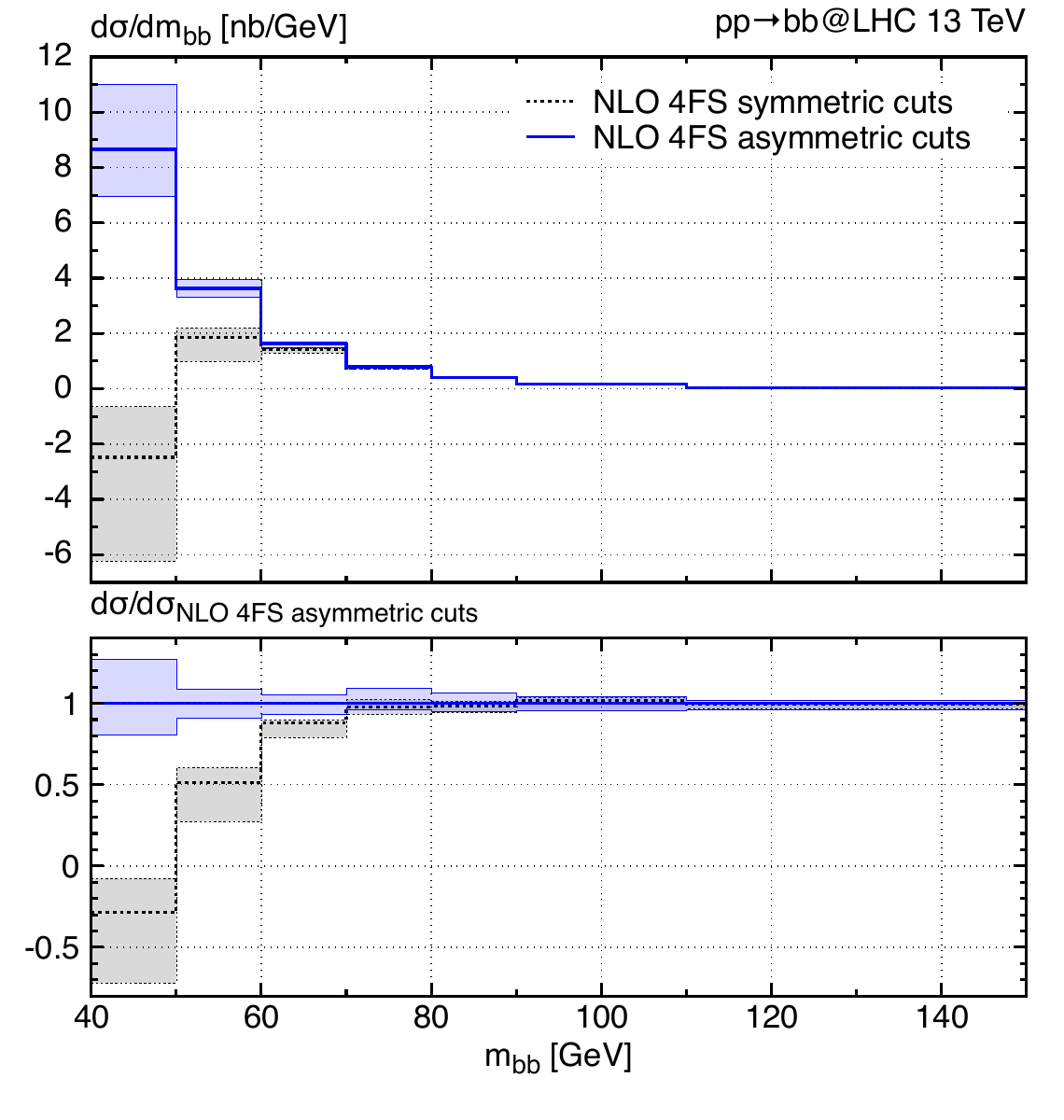}
&
\hspace{0.25cm}
\includegraphics[width=.42\textwidth,page=1]
{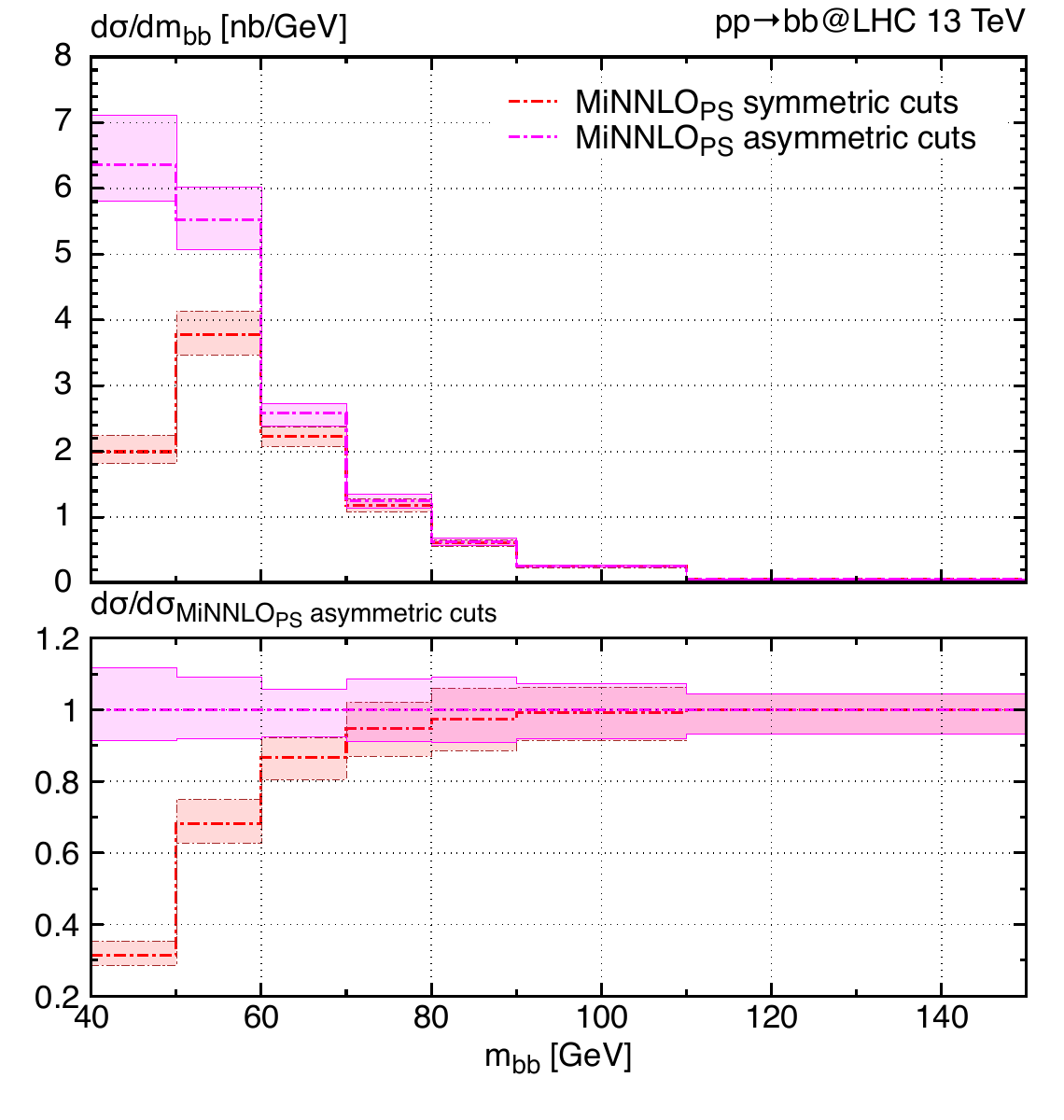}
\\
\end{tabular}
\vspace*{1ex}
\caption{
\label{fig:LHCbAsym}
Left: Prediction for the invariant mass spectrum of the $bb$ dijet system at NLO QCD accuracy when either symmetric or asymmetric selection cuts are applied to the transverse momentum of the jets, see text for details.
Right: Same as left panel, now with the \minnlo{} prediction.}
\end{center}
\end{figure}

The NLO QCD predictions for the invariant mass of the dijet $bb$-system is shown in the left panel of \fig{fig:LHCbAsym}.
We observe that, in the case of symmetric cuts, the fixed-order predictions for $m_{b\bar b}$ become negative in the region of $m_{b\bar b} \sim 40$\,~GeV, i.e.\ in the vicinity of $m_{b\bar b} \sim 2 p_{T,j}^{\rm min}$.
In the case of the asymmetric cuts, we do not observe such `pathological' behaviour.
The asymmetric cuts allow for contributions that involve a pseudo-soft gluon emission that lead to an imbalance in the $p_T$ of the two $b$-jets, which near threshold may be rejected by the symmetric cut selections.
In the right hand panel of \fig{fig:LHCbAsym} the same selections are considered with the \minnlo{} prediction.
This prediction, as a result of the resummation effects introduced by the parton shower, does not have the same pathological behaviour as fixed-order predictions.
However, the cross section in the region of $m_{b\bar b} \sim 40$\,~GeV changes by a factor of 2-3 when allowing for subleading $b$-jets with a lower threshold of $p_{T,j}^{\rm min} = 17$\,~GeV instead of 20\,~GeV.
The use of asymmetric cuts has been adopted for dijet measurements in Run I~\cite{CMS:2012ftr,ATLAS:2013jmu}, and we believe such an approach should also be considered for future measurements heavy-flavour jet measurements at LHCb such that the infrared sensitivity is reduced and fixed-order predictions can still be applied to interpret the data.
More generally, the use of symmetric cuts can also have implications for other measurements (such as on-/off-shell Drell-Yan etc.), see the discussions in \citeres{Salam:2021tbm,Buonocore:2021tke,Alekhin:2024mrq}, and it may also be interesting to consider their role given the forward (and asymmetric) kinematic acceptance of the LHCb detector.

\section{Mass effects and predictions for double tagged jets}\label{sec:add}

The 4FS \minnlo{} predictions presented so far include an exact treatment of bottom-quark mass effects up to NNLO QCD accuracy. 
The resulting predictions contain a range of different bottom-quark mass effects which we can be broadly categorised into those which contain: logarithmically enhanced bottom-quark mass effects, power corrections (potentially also logarithmically enhanced), as well as `constant' contributions that do not vanish in the limit $m_b\to0$.
In contrast, a massless approach (e.g.\ a 5FS approach) does not include power corrections in the bottom-quark mass but does include (typically resummed) logarithmically enhanced bottom-quark mass effects and those contributions that do not vanish in the limit $m_b\to0$.

To assess the importance of retaining the full bottom-quark mass dependence in our 4FS NNLO+PS calculation we take the following two approaches.
Firstly, we perform a comparison of our 4FS NNLO+PS calculation with those obtained with lower-order NLO+PS results obtained in both the massive four-flavour scheme and the massless five-flavour scheme.
This comparison highlights the importance of NNLO perturbative corrections in the massive scheme, as well as the impact of resummation effects in the bottom-quark mass.
These predictions are generated with \textsc{MG5\_aMC@NLO}~\cite{Alwall:2014hca} at NLO+PS using bottom-quark pair production in the 4FS as well as dijet production in the 5FS. 
Secondly, we also quantify the impact of power corrections in the bottom-quark mass for $b$-jet observables at $\mathcal{O}(\alpha_s^3)$ fixed-order accuracy.
For both comparisons we consider the inclusive $b$-jet transverse-momentum spectrum ($p_{T,b\text{-jet}}$), corresponding to the setup of the ATLAS measurement at $7$\,TeV in \citere{ATLAS:2011ac}. 

The first comparison is achieved in a straight-forward way by comparing the various (N)NLO+PS predictions, while the second comparison (which requires us to isolate the contributions to the cross section arising from power corrections in the bottom-quark mass) is more involved.
For this reason we have included in \app{app:PC} the technical details of how that comparison is performed, and provide just a brief summary of how this is achieved in what follows.
The extraction of power corrections in the bottom-quark mass is achieved after recognising that at small $m_b$ we can write the 4FS cross section as an asymptotic expansion in $m_b$
\begin{align}
\label{eq:4FS}
{\rm d}\sigma_{\rm 4FS}
&=
{\rm d}\sigma_{L}
+
{\rm d}\sigma_{0}
+
{\rm d}\sigma_{\rm PC}\,.
\end{align}
Here, ${\rm d}\sigma_{L}$ is understood to contain the terms logarithmically enhanced in
the limit $m_b\to0$, with logarithms of the form
$L=\ln(Q^2/m_b^2)$, where $Q$ denotes a characteristic hard scale.
The term ${\rm d}\sigma_{0}$ collects the contributions that remain
finite and non-vanishing as $m_b\to0$, while
${\rm d}\sigma_{\rm PC}$ contains terms suppressed by positive powers
of $m_b/Q$. 
Accordingly, 
\begin{align}\label{eq:mb0}
\left.{\rm d}\sigma_{\rm 4FS}\right|_{m_b\to0}
&=
{\rm d}\sigma_{L}
+
{\rm d}\sigma_{0}
+
\mathcal{O}\!\left(\frac{m_b^2}{Q^2}\right).
\end{align}
We note that the prediction obtained in the massless limit is often referred to as the `leading power' (LP) prediction, so we introduce the short-hand notation
\begin{align} \label{eq:LP}
{\rm d}\sigma_{LP}&= 
{\rm d}\sigma_{L}
+
{\rm d}\sigma_{0}\,.
\end{align}

The logarithmic terms receive contributions from several sources. For
initial-state collinear configurations, they can be obtained from the
perturbative expansion of the bottom-quark parton distribution. For
final-state configurations, the corresponding logarithms are related
to the perturbative fragmentation of partons into massive bottom
quarks. 
Additional logarithmic terms arise from converting the strong
coupling and the parton distributions between the four- and
five-flavour schemes. 
The constant term ${\rm d}\sigma_0$ can be obtained by direct calculation using standard methods for the treatment of IRC singularities related to the presence of massless quark, or numerically in the $m_b\to0$ limit, see \app{app:PC} for more details.

The power-suppressed contribution then follows from
\begin{align} 
\label{eq:PC}
{\rm d}\sigma_{\rm PC}
&= 
{\rm d}\sigma_{\rm 4FS}
-
{\rm d}\sigma_{LP}
\end{align}

For comparison, the corresponding 5FS result can be represented
schematically as
\begin{align}
\label{eq:5FS}
{\rm d}\sigma_{\rm 5FS}
&=
{\rm d}\sigma_{L}^{\rm res}
+
{\rm d}\sigma_{0}\,,
\end{align}
where ${\rm d}\sigma_{L}^{\rm res}$ denotes the logarithmic
contributions resummed through the bottom-quark parton distribution,
together with the corresponding five-flavour running of the strong
coupling. A consistently matched prediction would combine this
resummation with the finite-mass information contained in the 4FS
calculation while avoiding double counting of their common massless
limit. Schematically, this amounts to supplementing the 5FS result with
the power corrections extracted from the 4FS calculation, after the necessary perturbative and scheme conversions. 
A study of the size and behaviour of ${\rm d}\sigma_{\rm PC}$ can be useful to indicate the kinematic regions in which finite bottom-mass effects are relevant, or when they can be neglected from the predictions.

\begin{figure}[t!]
\begin{center}
\begin{tabular}{cc}
\includegraphics[width=.42\textwidth]
{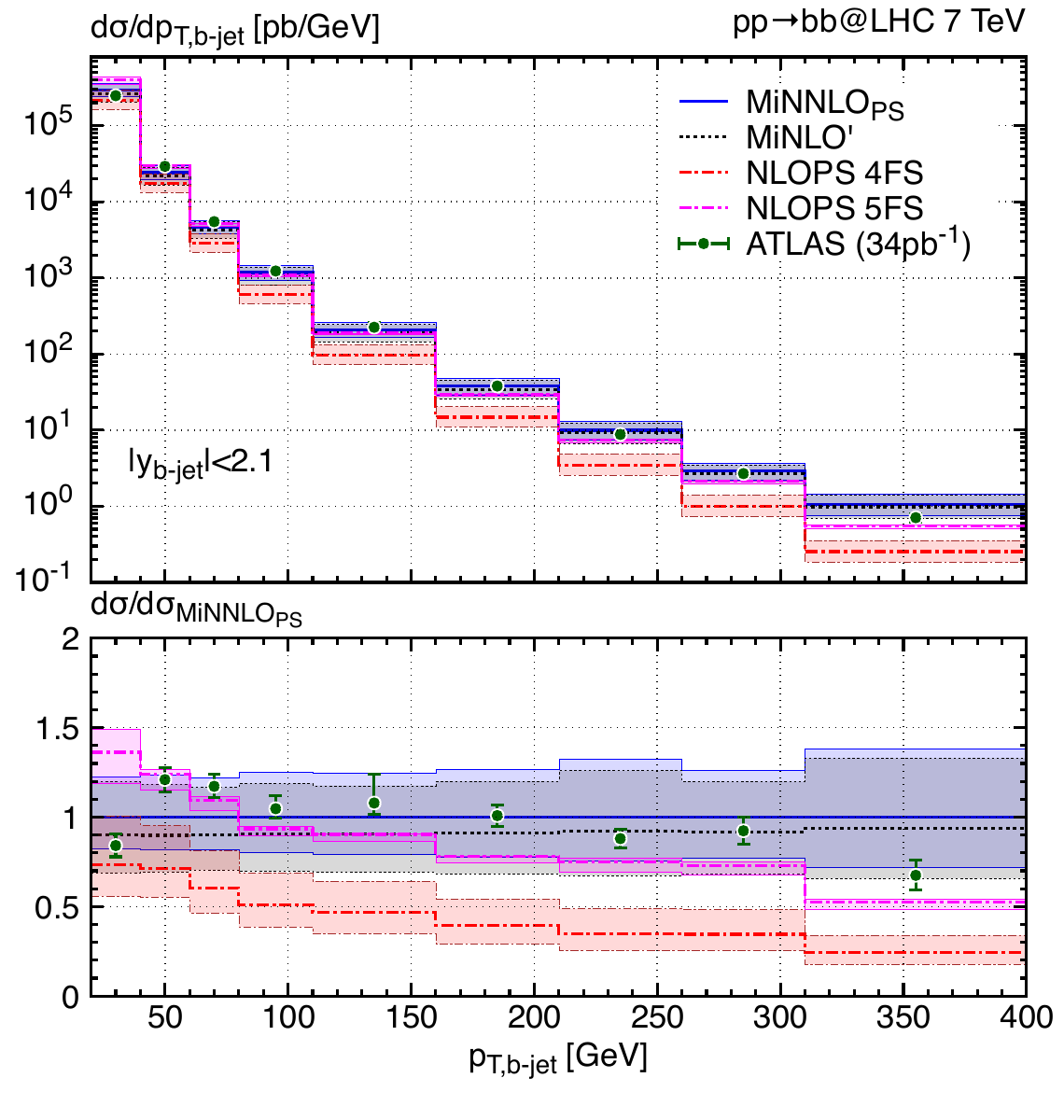}
&
\hspace{0.25cm}
\includegraphics[width=.42\textwidth,page=1]
{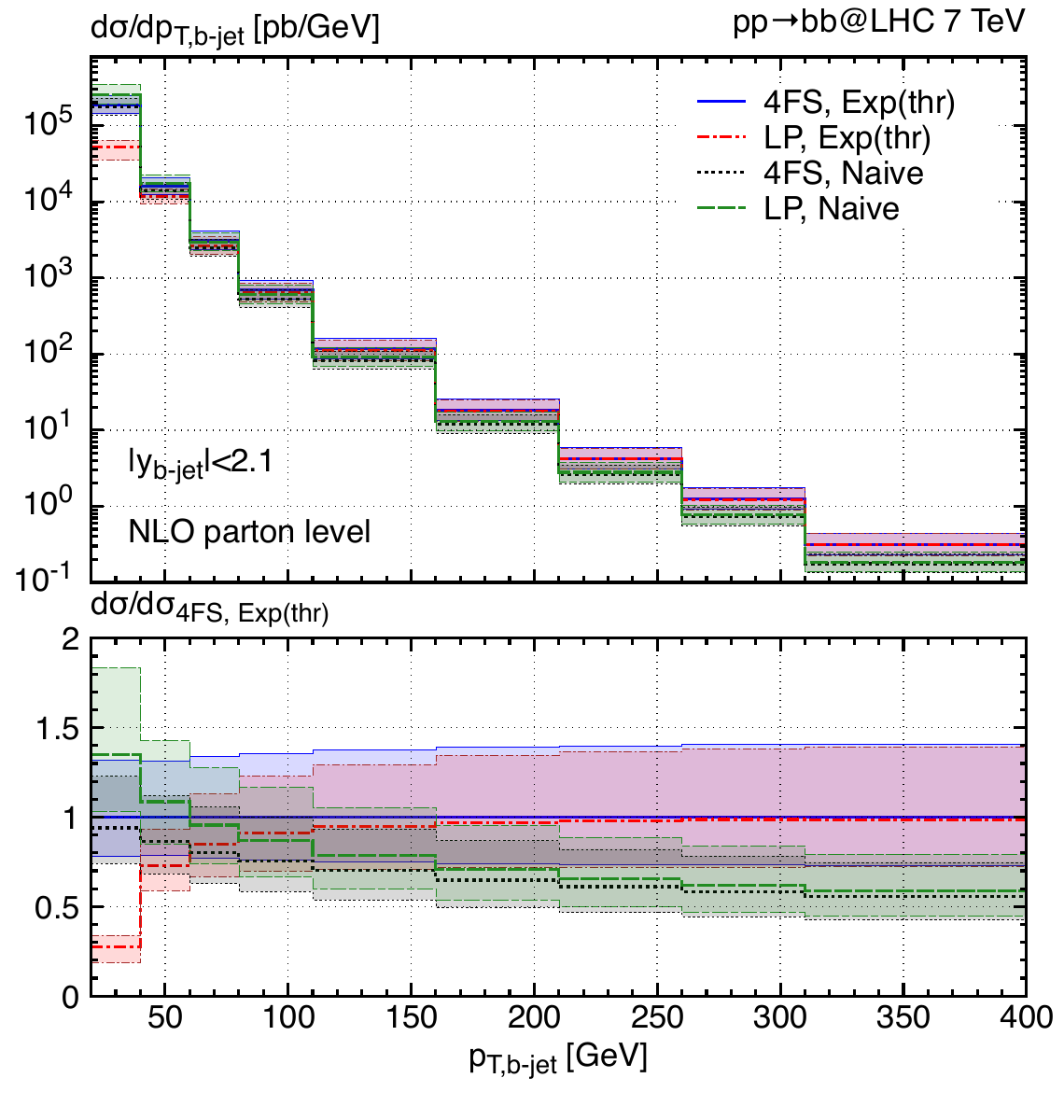}
\\
\end{tabular}
\vspace*{1ex}
\caption{
\label{fig:pc}
Left: Comparison of massive and massless predictions with the ATLAS
measurement of the inclusive $b$-jet transverse-momentum spectrum at
$7$\,TeV from \citere{ATLAS:2011ac}.
Right: Comparison of 4FS and the corresponding leading power (LP) predictions for the same distribution, where results with either {\tt exp(thr)} or {\tt naive} $b$-jet flavour definitions are shown.
See the text for details.
}
\end{center}
\end{figure}

With this in mind, we now turn to the comparison of the various (N)NLO+PS accurate predictions for the inclusive $b$-jet transverse-momentum spectrum ($p_{T,b\text{-jet}}$).
In all cases, the predictions are performed at hadron-level (with stable $B$-hadrons) and the impact of MPI is included.
%
The left panel of \fig{fig:pc} compares the ATLAS measurement (green data points) with the
\minnlo{} prediction (blue, solid histograms), the conventional 4FS
NLO+PS result (magenta, dash-dotted histograms), and the massless 5FS
NLO+PS result based on dijet production (red, dashed histograms).

The conventional 4FS NLO+PS prediction is insufficient to describe the
measurement, with deviations that substantially exceed its perturbative
scale uncertainties. The sizeable NNLO corrections included in
\minnlo{} are therefore essential for obtaining the good agreement
with the data observed in the previous section. It is also instructive
to recall that the formally NLO-accurate \minlo{} prediction considered
there already provides a substantially better description than the
conventional NLO+PS result. This reflects the additional higher-order
contributions incorporated through the \minlo{} construction.

The massless 5FS NLO+PS calculation, by contrast, exhibits remarkably
small scale uncertainties. These appear unnaturally narrow for an
NLO-accurate prediction and are unlikely to provide a realistic
estimate of the missing higher-order corrections. In particular, the
5FS uncertainty band is substantially smaller than that of the
NNLO-accurate \minnlo{} prediction. At the level of the central values,
the 5FS result nevertheless describes most of the measured spectrum
reasonably well. In several bins, it is even closer (or as close) to the data than
the central \minnlo{} prediction, and it reproduces the shape of the
high-$p_{T,b\text{-jet}}$ tail quite well. Owing to its very small uncertainty
band, however, it is incompatible with the data within uncertainties
in almost every bin. Moreover, it substantially overestimates the
cross section in the lowest-$p_{T,b\text{-jet}}$ bin, which contains
the bulk of the measured rate.

Since finite-mass effects are generally expected to be largest at small
$p_{T,b\text{-jet}}$, one might attribute the poor description of the
first bin by the 5FS calculation to the absence of power corrections.
This motivates the explicit extraction of the power corrections as outlined above.
To achieve this, we perform the leading-power calculation for $b$-jet production up to $\mathcal{O}(\alpha_s^3)$ as introduced in \eqn{eq:LP}. 
A detailed discussion on this calculation in the context of different flavour tagging approaches can be found in \app{app:PC}.
In \fig{fig:pc} we directly present the results of this calculation, where 4FS and leading-power (labelled as LP) predictions for the inclusive $p_{T,b\text{-jet}}$ distribution are compared. 
Each of the predictions are shown for the {\tt naive} as well as the experimental tagging approach {\tt exp(thr)} which requires the presence of $p_T$ threshold on the tagging bottom quark, and the impact of scale variations are included for all predictions. 
The impact/size of the power corrections in $m_b$, see \eqn{eq:PC}, is simply the difference between the 4FS and LP predictions.

Focussing on {\tt exp(thr)} tagging predictions in the small $p_{T,b\text{-jet}}$ region, the difference of the 4FS (blue, solid) and the LP (red, dash dotted) predictions is a large and positive ($\approx +60\%$) correction.
Consequently, adding these mass corrections to the 5FS calculation (together constituting a matched prediction) would therefore further increase the cross section and substantially worsen the discrepancy in this region. 
This observation also reinforces the conclusion that the very small scale uncertainties
of the 5FS NLO+PS result (as shown in the right panel of \fig{fig:pc}) do not provide a reliable estimate of its theoretical uncertainty.

The right panel of \fig{fig:pc} also illustrates two general features of
the finite-mass corrections. First, as expected, they are largest at
small $p_{T,b\text{-jet}}$ and decrease towards the tail of the
distribution. An accurate massive calculation, such as the \minnlo{}
generator presented here, is therefore particularly important in the
low-$p_{T,b\text{-jet}}$ region. 
Second, both the magnitude and the sign
of the power corrections depend strongly on the jet-flavour definition.
This can be observed by comparing the difference of the 4FS (black, dotted) and LP (green, dashed) predictions in the {\tt naive} case, which indicate that power corrections are negative.
Hence, the power corrections obtained with the experimental and
{\tt naive} tagging definitions have opposite signs. This demonstrates that
finite-mass effects and flavour assignment cannot be considered
independently: the choice of jet-flavour definition affects not only
the infrared structure of the prediction, but also the numerical size
of the mass-suppressed contributions.

In the longer term, a consistent combination of the NNLO-accurate
massive 4FS calculation with the resummation of collinear logarithms
provided by the 5FS would offer the most complete theoretical
description. Such a matched calculation could retain the finite-mass
effects that are important at low transverse momentum while exploiting
the logarithmic resummation of the 5FS at large scales, see for example~\citere{Hoche:2019ncc} for work in this direction. Such an approach will become increasingly important in view of the larger data sets and reduced
experimental uncertainties expected at the High-Luminosity LHC.

Before summarising, we wish to comment further on the behaviour of the power corrections in the {\tt exp(thr)} and {\tt naive} tagging approaches, where opposite signs for the corrections are observed for the two cases.
We have confirmed that, for this observable, the power corrections for both {\tt exp(thr)} and {\tt exp} tagging behave in a very similar way.
This indicates that observed difference for the sign of the power corrections in the {\tt naive} case relates to the exclusion of double-tagged $b$-jets.

Double-tagged $b$-jets naturally arise in the {\tt exp} tagging procedure, where it is required that a $b$-jet contains \textbf{one or more}  $B$-hadrons (bottom quarks).
In the massive NLO calculation, a double-tagged contribution can originate from a configuration involving a pseudo-collinear splitting of the form $g\to b\bar b$, meaning that the cross-section involving a double-tagged jet is logarithmically enhanced for small values of the angular separation $\Delta R_{bb}$ between the quark pair.
Such jets form a QCD background for many SM processes (e.g.\ those involving hadronic Higgs boson decay to bottom quarks), and their study is therefore one of considerable interest~\cite{ATLAS-CONF-2016-039,ATLAS-CONF-2016-002,CMS-PAS-BTV-13-001,Ilten:2017rbd,ATLAS:2018zhf,LHCb:2025tvf}.
Moreover, an understanding of such jets (both experimentally and in simulation) is crucial in order to apply IRC-safe jet-flavour algorithms~\cite{Banfi:2006hf,Czakon:2022wam,Gauld:2022lem,Caola:2023wpj} either in measurement, 
or in the process of unfolding~\cite{Gauld:2020deh,ATLAS:2024tnr}.

In this context it is interesting to study how the 4FS \minnlo{} prediction compares to the various NLO+PS predictions. 
To achieve this we consider the rate of double-tagged $b$-jets reconstructed with the anti-$k_T$ algorithm with a fat $R = 1.0$ jet obtained at 7~TeV collisions.
To focus on the role of perturbative corrections, we have turned off the effects of hadronization and MPI.

\begin{figure}[t!]
\begin{center}
\begin{tabular}{cc}
\includegraphics[width=.42\textwidth]
{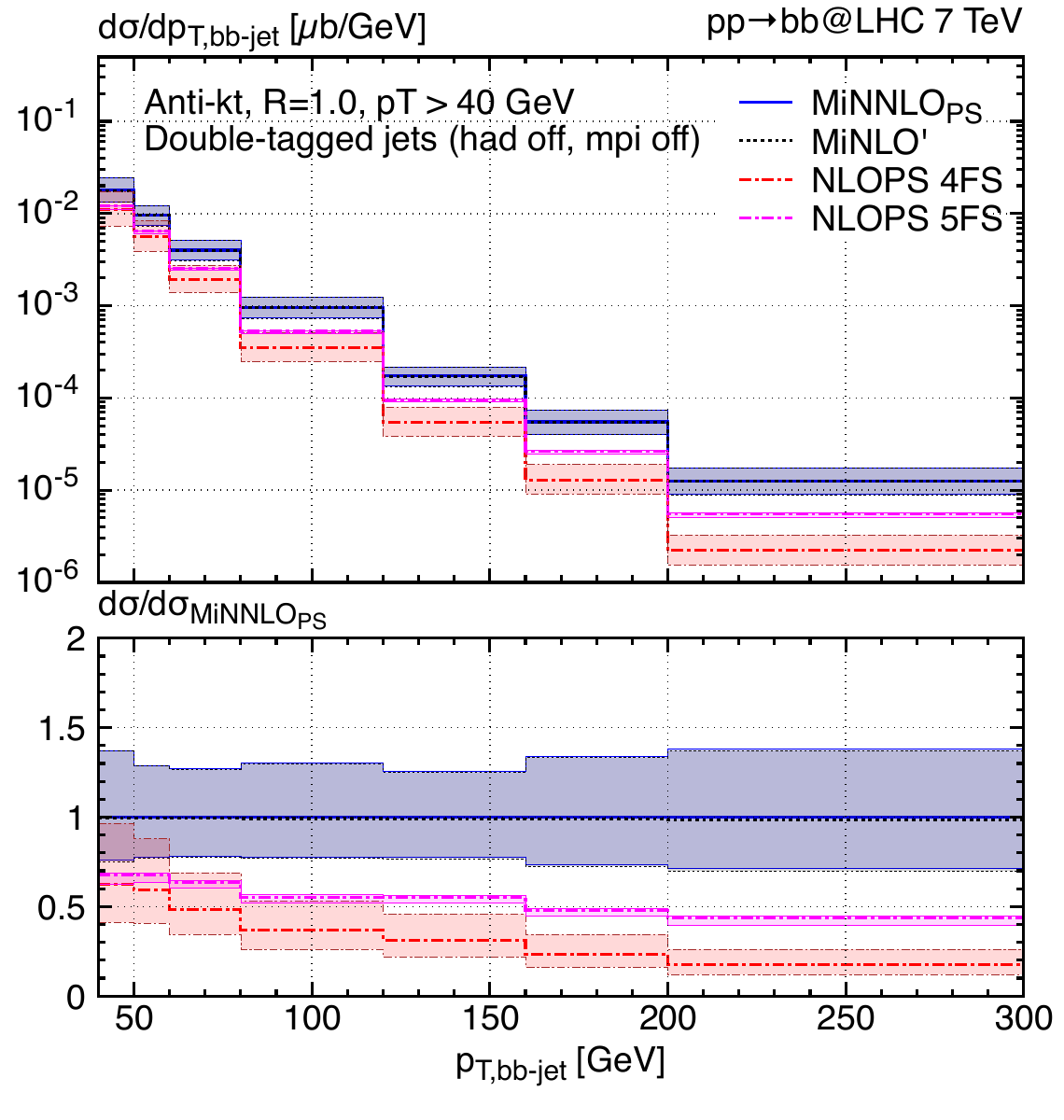}
&
\hspace{0.25cm}
\includegraphics[width=.42\textwidth,page=1]
{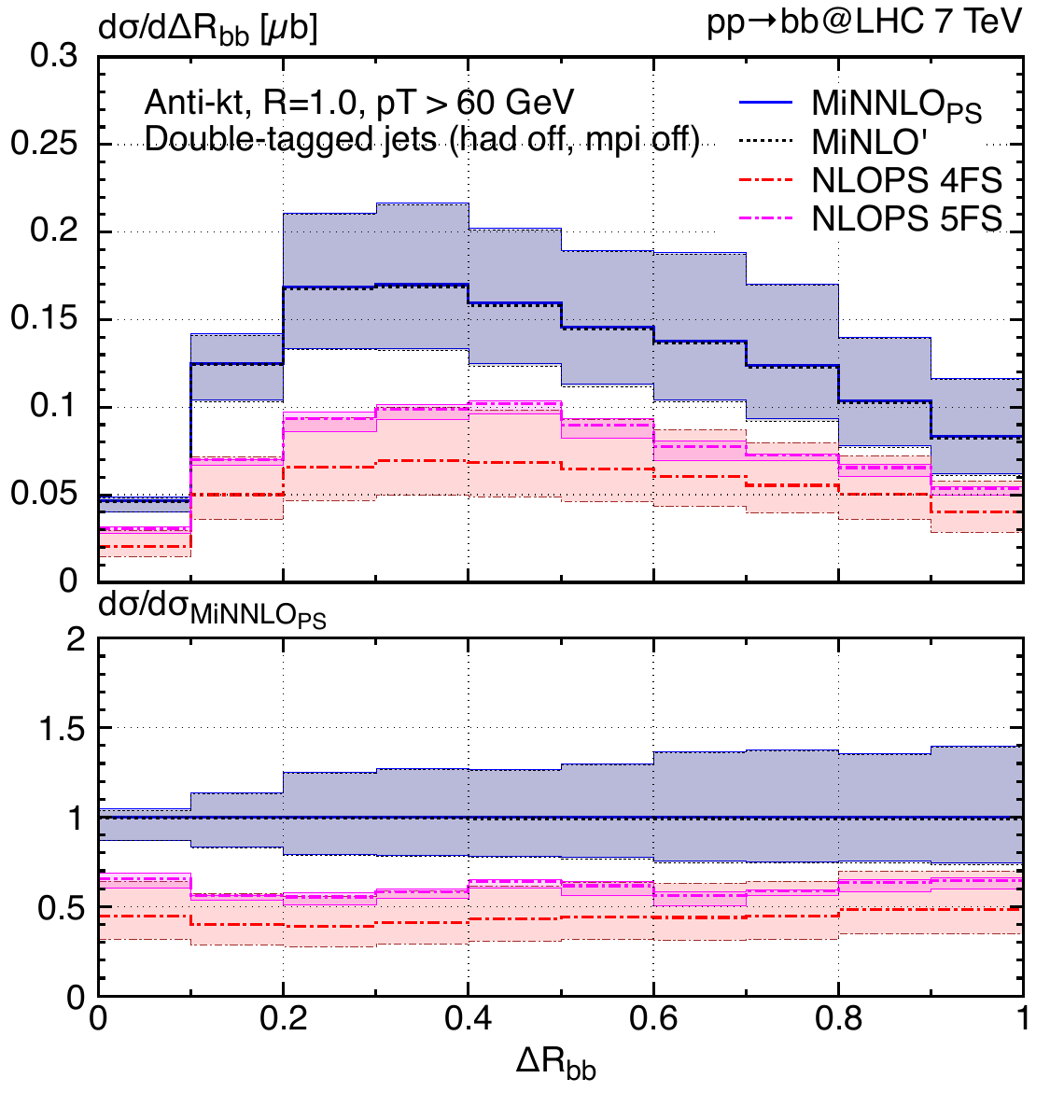}
\\
\end{tabular}
\vspace*{1ex}
\caption{
\label{fig:double}
Left: Transverse momentum distribution of $R=1.0$ double tagged anti-$k_T$ $b$-jets. Right: Distribution of distance separation $\Delta R_{bb}$ between bottom-quark pairs within double tagged jets with $p_{T} > $\,60~GeV.}
\end{center}
\end{figure}

The results are shown in \fig{fig:double} where in the left panel we consider inclusive transverse-momentum spectrum of double-tagged $b$-jets ($p_{T,{bb\text{-jet}}}$), and in the right panel we consider the distance separation $\Delta R_{bb}$ between the constituent bottom and anti-bottom quark of the double-tagged jet when requiring $p_{T,{bb\text{-jet}}} > $\,60~GeV.

For the $p_{T,{bb\text{-jet}}}$ distribution, it is observed that the \minnlo{} prediction receives substantial corrections  with respect to the 4FS NLO+PS prediction, and that this correction grows with increasing $p_{T,{bb\text{-jet}}}$.
The 5FS NLO+PS prediction remains roughly a factor of two smaller for the considered $p_{T,{bb\text{-jet}}}$ range.
We note that the presence of such a large correction is not unexpected, as the double tag distribution is effectively a LO observable for the 4FS NLO+PS prediction.
It is also for this reason that \minlo{} and \minnlo{} predictions are effectively the same.

Focussing on the $\Delta R_{bb}$ distribution in the right panel of \fig{fig:double}, which is obtained with \mbox{$p_{T} > 60$\,GeV}, we observe that the shape is described similarly for all predictions. In terms of normalisation, the \minnlo{} prediction receives a large correction (+50\%) compared to the 4FS NLO+PS prediction.
Although not shown here, a similar behaviour was also observed when hadronization was activated.

Previous experimental studies from the ATLAS collaboration at 13\,TeV focused on the measurement of double-tagged $b$-jets \cite{ATLAS:2018zhf}.
However, they conclude that the theoretical prescription with current tools is not particularly good and that improvements are required. 
Therefore, it would be interesting to extend these studies with our state-of-the-art NNLO+PS simulation in the massive scheme.
Furthermore, given the differences in shape and normalization for the various predictions reported above,
it would be important to study experimental measurements of absolute cross sections of double-tagged $b$-jet distributions (not just normalized ones).

\section{Summary}

To summarize, we have presented new NNLO+PS predictions for $b$-jet production at the Large Hadron Collider. To this end, we employed the \minnlo{} generator for bottom-quark pair production and interfaced it with the \PYTHIA{8} parton shower, including hadronization and multi-parton-interaction effects, to obtain a realistic description of $b$-jet observables.

Using this setup, we carried out a phenomenological comparison of recently proposed jet-flavour algorithms \cite{Gauld:2022lem,Caola:2023wpj}. For the process and observables considered in this work, where bottom quarks are treated as massive, we find no significant differences between the various definitions beyond the {\tt naive} treatment of $g\to b\bar b$ splittings. 
In particular, none of the new algorithms leads to a noticeable reduction of perturbative uncertainties for the observables considered here. 
Likewise, we observe no appreciable differences between infrared-safe flavour definitions (which avoid logarithms of the bottom-quark mass in the four-flavour scheme) and a {\tt naive} flavour definition (using the standard experimental anti-$k_T$ definition while not considering a $b\bar b$ jet as a $b$-jet).

In the second part of our phenomenological study, we compared the
\minlo{} and \minnlo{} predictions with recent LHC measurements of
$b$-jet production using the same anti-$k_t$ jet-clustering procedure
employed in the experimental analyses. We found that \minnlo{} yields
sizeable NNLO QCD corrections with respect to \minlo{}, reaching up to
$30\%$ for resolved two-$b$-jet observables, such as the invariant mass
of the two leading $b$-jets, whereas inclusive $b$-jet observables,
such as the inclusive $b$-jet transverse-momentum spectrum, receive
smaller corrections of approximately $10\%$. 
At the same time, \minnlo{} reduces the perturbative scale
uncertainties significantly in kinematic regions close to the back-to-back limit of two $b$-jets.

Overall, we find excellent agreement with the 7\,TeV
measurements by ATLAS \cite{ATLAS:2011ac,ATLAS:2016anw} and CMS
\cite{CMS:2012pgw}. By contrast, the recent 13\,TeV LHCb measurement
\cite{LHCb:2020frr} exhibits a significant deficit in the measured cross
section with respect to both the \minlo{} and \minnlo{} predictions.
While this discrepancy is already present at NLO accuracy, the positive
NNLO QCD corrections and the reduced theoretical uncertainties make it
more pronounced. Interestingly, rescaling the LHCb data by a factor of
two leads to excellent agreement with the \minnlo{} prediction for both
the inclusive cross section and the differential distributions.
Nevertheless, future measurements with increased statistics, together
with independent measurements by ATLAS and CMS at $13$\,TeV,
will be essential to clarify the origin of this discrepancy.

We also investigated the role of finite bottom-quark mass effects by
extracting the corresponding power corrections and comparing massive
4FS and massless 5FS NLO+PS predictions. We find sizeable power corrections,
particularly at low transverse momenta, whose magnitude and sign depend
sensitively on the jet-flavour definition. Moreover, the perturbative
uncertainties of conventional massless 5FS NLO+PS predictions appear
unrealistically small and fail to cover the observed discrepancies with
the measurements.

We look forward to future measurements of $b$-jet production with the available and expected LHC data sets.
We expect the \minnlo{} $b\bar b$ generator and the corresponding NNLO+PS predictions for $b$-jet production to provide an important tool for precision studies of heavy-flavour jet production at the LHC.

In the longer term, a consistent combination of the NNLO-accurate massive 4FS calculation with the resummation of collinear logarithms provided by the 5FS in the context of \minnlo{} represents a natural direction towards further improving the theoretical precision of heavy-flavour jet predictions.

\noindent {\bf Acknowledgements.}
We are indebted to Javier Mazzitelli for several discussions in the beginning of the project.
We have used the Max Planck Computing and Data Facility (MPCDF) in
Garching to carry out all simulations presented here.

\appendix
\section{Extraction of power corrections} \label{app:PC}
In \sct{sec:add}, the 4FS \minnlo{} predictions have been compared to NLO+PS predictions obtained in either the 4FS or 5FS, where we find the predictions to exhibit different behaviour in the $b$-jet transverse momentum.
To better understand the differences between these predictions we have performed a quantitative study of quark mass effects at fixed-order.
The main results of that study are shown in the right plot of \fig{fig:pc}, with a corresponding accompanying discussion. 
As the method to extract these power corrections can be useful for other jet-based studies, we sketch the main steps of the procedure in what follows.

A procedure for the extraction of bottom-quark mass effects for inclusive $b$-jet production at $\mathcal{O}(\alpha_s^3)$ (i.e. beyond the first trivial order) has been discussed in \citere{Banfi:2007gu}. This procedure has also been extended to the case of $Z+b$-jet production up to $\mathcal{O}(\alpha_s^3)$ in~\citere{Gauld:2020deh} where the flavour-$k_{T}$ algorithm~\citere{Banfi:2006hf} was applied---effectively requiring an NLO extraction of mass effects as they are already non-trivial at $\mathcal{O}(\alpha_s^2)$.
A similar procedure has been applied to the case of (on-shell) $Z+b$-jet in combination with a variant of the {\tt naive} tagging in~\citere{Generet:2025gdy}.
In short, this method can be applied to differential observables provided that the massless limit of the calculation can be predicted.
According to the discussion in \sct{sec:add}, the massless limit of the differential cross section can be decomposed into those terms which contain a logarithmic dependence on the bottom-quark mass (${\rm d}\sigma_{L}$) and a remaining constant (${\rm d}\sigma_{0}$).
Knowledge of both ${\rm d}\sigma_{L}$ and ${\rm d}\sigma_{0}$ is required for a numerical extraction of the component of the cross section due to power corrections, see \eqn{eq:PC}.
However, as discussed in Appendix A of \citere{Banfi:2007gu}, it is also possible to numerically extract ${\rm d}\sigma_{0}$ when ${\rm d}\sigma_{L}$ is known.
Introducing the notation
\begin{align}
{\rm d}\sigma_{M-L}
\equiv
{\rm d}\sigma_{\rm 4FS} - {\rm d}\sigma_{L}\,,
\end{align}
it is then clear from rearrangement of \eqn{eq:mb0} that
\begin{align}\label{eq:4fsmL}
{\rm d}\sigma_{M-L} 
\underset{m_b \to 0}{=}
{\rm d}\sigma_{0}
+
\mathcal{O}\!\left(\frac{m_b^2}{Q^2}\right).
\end{align}
Thus, the constant can be extracted by numerically evaluating the left-hand side of \eqn{eq:4fsmL} for decreasing values of $m_b$.
In this work, depending on the considered definition of the jet flavour, we obtain ${\rm d}\sigma_{0}$ either by direct calculation or numerically in the $m_b\to0$ limit.
These two cases are distinguished below.
In the following we describe the procedure to extract the power corrections for $b$-jet production from the 4FS NLO calculation, that is up to $\mathcal{O}(\alpha_s^3)$.

\subsection{Naive tagging}
When an IRC-safe jet-flavour algorithm is applied, the logarithmic dependence of the cross section ${\rm d}\sigma_{L}$ can be constructed by making use of a combination of the heavy-flavour decoupling relations for PDFs and the strong coupling (see also the discussion in Section 2 and 3 of \citere{Gauld:2021zmq}). 
For the NLO calculation in the 4FS with the {\tt naive} tagging procedure (which tags/labels a jet containing a $b \bar b$ pair as flavourless) the logarithmic dependence can be obtained following the same procedure.
Below, we sketch how the logarithmic cross section arising from the various partonic channels can be constructed up to $\mathcal{O}(\alpha_s^3)$ for the $b$-jet production process.
Working at the partonic level, and distinguishing between initial-state channels in the 4FS (real-emission processes: $q\bar{q}\to b\bar b g$, $q g\to b\bar b q$, $gg\to b\bar b g$ with massive bottom quarks), we can introduce the expressions for the logarithmic cross section of the form:
\begin{align} \label{eq:naiveL} \nonumber
{\rm d} \hat{\sigma}^{q\bar q,{\rm naive}}_{L} &\sim  2 \,\Delta \alpha_s(\mu_R/m_b)  {\rm d} \hat{\sigma}_{q\bar q\to b\bar b}^{m_b=0} \,, \\ \nonumber
{\rm d} \hat{\sigma}^{qg,{\rm naive}}_{L} &\sim  \hat{A}_{g\to b}(z,\mu_F/m_b) \otimes \left( {\rm d} \hat{\sigma}_{qb\to qb}^{m_b=0} + {\rm d} \hat{\sigma}_{q\bar b\to q\bar b}^{m_b=0} \right) \,, \\ \nonumber
{\rm d} \hat{\sigma}^{gg,{\rm naive}}_{L} &\sim  2 \left( \Delta \alpha_s(\mu_R/m_b) + \hat{A}_{g\to g}(z,\mu_F/m_b) \otimes \right)  {\rm d} \hat{\sigma}_{gg\to b\bar b}^{m_b=0} \\ &\quad + 2 \, \hat{A}_{g\to b}(z,\mu_F/m_b) \otimes  \left( {\rm d}\hat{\sigma}_{bg\to bg}^{m_b=0} + {\rm d} \hat{\sigma}_{\bar b g\to \bar b g}^{m_b=0}\right) \,,
\end{align}
where ${\rm d} \hat{\sigma}_{ij\to kl}^{m_b=0}$ denotes the tree-level partonic cross section from the appropriate underlying $2\to2$ process evaluated with $m_b = 0$ i.e.\ in the 5FS. 
In the first line of \eqn{eq:naiveL}, the logarithmic approximation is proportional to $\Delta \alpha_s$, which describes the scheme change from the decoupling to the $\MSbar$ treatment of the heavy-flavour content in the $\alpha_s$ renormalisation constant.
In the second line of \eqn{eq:naiveL}, the logarithmic approximation for the full $q g \to b\bar b q$ subprocess receives contributions from the PDF decoupling relation, and one should include similar corrections for the $\bar q g$, $g q$, and $g \bar q$ channels.
In the last line, the gluon-fusion channel receives contributions from both PDF and $\alpha_s$ decoupling relations.
Expressions for the PDF decoupling relations up to $\mathcal{O}(\alpha_s^2)$ can be found in the appendix of \citere{Buza:1996wv}, and those for $\alpha_s$ can be found in \citere{Chetyrkin:2000yt}. 
As they are compact, we quote the relevant $\mathcal{O}(\alpha_s)$ results for the $b$-quark here
\begin{align} \nonumber
\hat{A}_{g\to b}^{(1)}\left(z, \mu_F/m_b\right) &= \frac{\alpha_s}{2\pi} P^{(0)}_{g\to b}(z) \ln \left[\frac{\mu_F^2}{m_b^2} \right]   \,,\\ \nonumber
\hat{A}_{g\to g,b}^{(1)}\left(z, \mu_F/m_b\right) &=  \frac{\alpha_s}{2\pi} \delta(1-z) \left( - \frac{2 T_R}{3} \right)  \ln \left[\frac{\mu_F^2}{m_b^2} \right]   \,,\\
\Delta \alpha_s^{(1)}\left(\mu_R/m_b\right) &=  \frac{\alpha_s}{2\pi} \frac{2 T_R}{3}  \ln \left[\frac{\mu_R^2}{m_b^2} \right]  \,,
\end{align}
with $T_R = 1/2$, and $P^{(0)}_{g\to b}(z)$ the leading-order DGLAP splitting function for gluon splitting to a massless quark pair, and $\alpha_s \equiv \alpha_s(\mu_R)$.

It is understood that the hadronic cross section for each of the channels is obtained after convolution of the partonic cross sections with  the appropriate PDFs, integration over phase space, and application of the final-state selections.
The form of these corrections is the same for other IRC-safe jet-flavour algorithms, and can also be found in Eqs.\,(11) and (12) of \citere{Banfi:2007gu}.

The direct calculation of ${\rm d}\sigma_{0}$ for each of these channels can be performed with subtraction techniques applicable to massless quarks.
If the subtraction scheme used for the corresponding massive calculation has a smooth $m_b\to0$ limit (such as the massive dipole subtraction scheme used here), then the extension to the massless case is not too complicated.
In the case of the $q \bar q$ channel, after applying the appropriate change of scheme for the $\alpha_s$ renormalisation counter-term, the massless calculation is obtained by performing the calculation with $m_b = 0$.
For the $qg$- and $gg$-induced channels the massless constant can also be obtained from the corresponding massive calculation in a similar way. 
In those cases one must additionally account for (at real-subtraction and integrated-subtraction levels) the presence of unresolved initial-state collinear configurations involving the heavy-flavour quark.
In the massless calculation, the explicit divergence associated with this integrated subtraction term is removed by a corresponding mass factorisation counter-term.

\subsection{{\tt Exp} tagging}
In contrast to the {\tt naive} tagging approach, the {\tt exp} tagging procedure labels a jet as flavoured if it contains one or more bottom quarks (or $B$-hadrons), i.e.\ jets containing a $b \bar b$ pair are considered flavoured in this case.
Such a selection is not collinear safe in the $m_b\to 0$ limit and introduces new logarithmic corrections as compared to the {\tt naive} case at finite  $m_b$.
In the massive calculation, those corrections arise exactly from double-tagged jets which contain a (pseudo) collinear $b \bar b$.
With some additional considerations, it is also possible to construct both ${\rm d}\sigma_{0}$ and ${\rm d}\sigma_{L}$ in this case.

As an example, we consider the real emission subprocess $q\bar q \to g b\bar b$. In the pseudo-collinear limit of the $b \bar b$ pair, one can approximate the correction while retaining differential information on the collinear momentum fraction $z$ of the outgoing  bottom quark according to
\begin{align} \label{eq:gtobb}
{\rm d} \hat{\sigma}_{q\bar q\to gg}^{m_b=0} \otimes \left[ \frac{\alpha_s}{2\pi} \int_0^1 {\rm d}z \, P^{(0)}_{g\to b}(z)  \ln \left( \frac{\mu^2}{m_b^2}\right) \right]\,.
\end{align}
If one is inclusive in the momentum fraction $z$ (such is the case in the {\tt exp} tagging procedure) the $z$ integral can be performed analytically to obtain the correction
\begin{align}
{\rm d} \hat{\sigma}_{q\bar q\to gg}^{m_b=0} \left[ \frac{\alpha_s}{2\pi} \frac{2 T_R}{3}   \ln \left( \frac{\mu^2}{m_b^2}\right) \right]\,.
\end{align}
The term in square brackets is the same as that for the decoupling term $\Delta \alpha_s(\mu/m_b)$. The logarithmic approximation for the $q\bar q$ channel in the {\tt exp} tagging procedure therefore can be written as
\begin{align} \label{eq:expL} \nonumber
{\rm d} \hat{\sigma}^{q\bar q,{\rm exp}}_{L} &\sim {\rm d} \hat{\sigma}^{q\bar q,{\rm naive}}_{L}  +  2 \, \Delta \alpha_s(\mu/m_b) {\rm d} \hat{\sigma}_{q\bar q\to gg}^{m_b=0}\,.
\end{align}
Similar adaptations must also be made for the $qg$ and $gg$ channels.

A direct calculation of ${\rm d}\sigma_{0}$ can also be performed when the {\tt exp} tagging procedure is applied.
As compared to the {\tt naive} tagging approach, one must take care to account for the presence of the unresolved collinear splittings of the form $g\to b \bar b$ which the jet algorithm no longer rejects.
Viewed in another way, one can identify/extract the bottom-quark component from the full massless dijet calculation by studying the $n_f$ dependence of the calculation in the relevant partonic channels (i.e.\ the $qg$, $gg$, and $q \bar q$ channels where $q = u,d,s,c$).

\subsection{{\tt Exp(thr)} tagging}
Finally, we turn to the most complicated case of the {\tt exp(thr)} tagging, where a momentum selection (typically a $p_T$ threshold) is applied to the bottom quarks  (or $B$-hadrons).
The momentum selection on the bottom quark (or $B$-hadron) is collinear unsafe in the massless limit and introduces yet a new source of logarithmic correction compared with the {\tt naive} and {\tt exp} tagging cases.

To understand the origin of this sensitivity, it is useful to consider a simple configuration such as the production of two back-to-back $b$-jets. 
Focussing again on the $q\bar q$ channel for simplicity, the leading-order prediction for this configuration is obtained from the partonic cross section for the subprocess $q \bar q \to b \bar b$, where the $b$ and $\bar b$ quarks are produced back-to-back, satisfying both the jet selections and any bottom-quark threshold (since the $b$-jets here are described by one particle).
At the real emission level, the emission of a hard collinear gluon (with respect to either $b$ or $\bar b$ quarks) will change the momentum (and $p_T$) of the quark (but not of the jet), meaning that a jet tagging requirement based on bottom-quark (or $B$-hadron) momentum will introduce a collinear sensitivity to the cross section.

The logarithmic approximation for the cross section can also be constructed in this scenario. Focussing on the above example, one adjusts the partonic cross section to account for potential collinear emissions on either of the external $b$ or $\bar b$ legs, and introduces a selection criterion on the identified $b(\bar b)$ quark momenta.
For example, considering the $b$-quark,
\begin{align}
{\rm d} \hat{\sigma}_{q\bar q\to b\bar b}^{m_b=0} \otimes \left[ \frac{\alpha_s}{2\pi} \int_0^1 {\rm d}z \, P^{(0)}_{b\to b}(z)  \ln \left( \frac{\mu^2}{m_b^2}\right) \right] \Theta_{\tt thr}\left(z p_b, p_{T}^{\rm thr}\right) \,,
\end{align}
where schematically the theta function $\Theta_{\rm thr}$ applies a final state selection on the fragmented momentum $z p_b$ based on the threshold restriction 
$p_{T}^{\rm thr}$.
Here it is understood that the momenta of the reconstructed jets are not modified, but that the momentum of the identified particle (that enters the {\tt exp(thr)} tagging procedure) is modified.
%

Together with the correction for $g\to b\bar b$ splittings differential in the momentum fraction $z$ in \eqn{eq:gtobb}, one can construct the logarithmic cross section for the {\tt exp(thr)} tagging procedure in all channels.
In practice, one can implement the logarithmic cross section for the {\tt exp(thr)} case in its general form, and obtain the {\tt exp} and {\tt naive} as special cases, i.e.\ integrating over $z$ to obtain the  {\tt exp} case, and further removing jets that contain a $b\bar b$ to obtain the {\tt naive} case.

The direct calculation of the massless constant ${\rm d}\sigma_{0}$ for the {\tt exp(thr)} case requires the use of subtraction methods for identified particles.
We did not implement this calculation, but instead obtained the result in the massless limit according to \eqn{eq:4fsmL}.
To obtain the constant through direct calculation, one must take care to include the full heavy-quark fragmentation function~\cite{Mele:1990cw} for $b\to b$ transitions that includes the initial condition (i.e.\ an $m_b$-independent correction).

\subsection{Numerical validation of the procedure}
\begin{figure}[t!]
\begin{center}
\begin{tabular}{cc}
\includegraphics[width=.42\textwidth]
{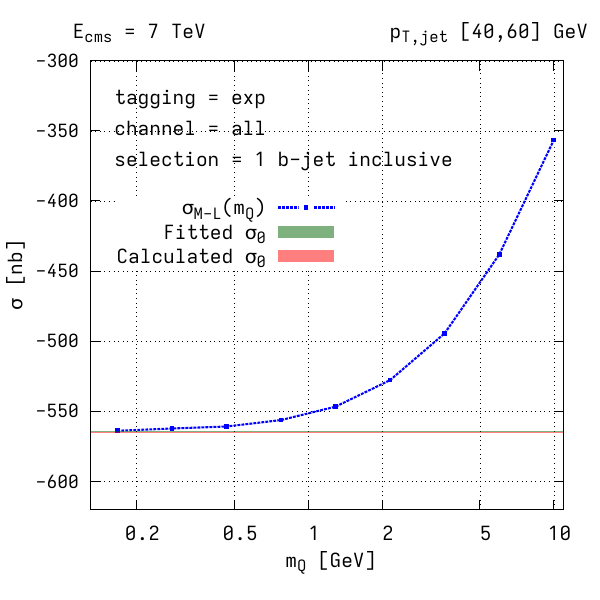}
&
\hspace{0.25cm}
\includegraphics[width=.42\textwidth]
{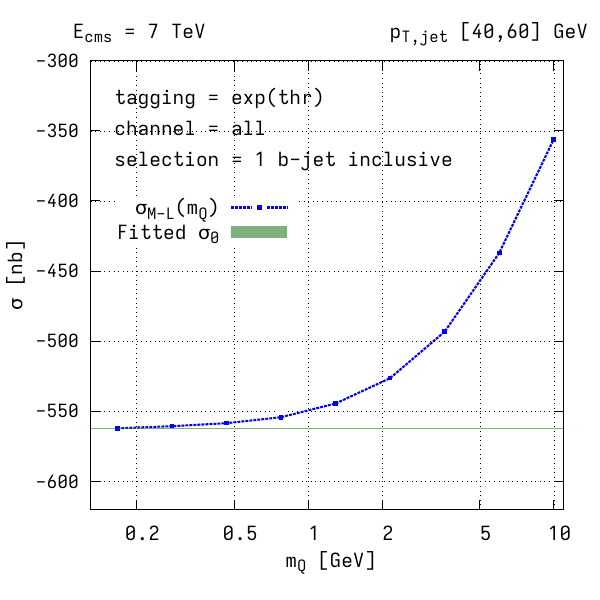}
\\
\end{tabular}
\vspace*{1ex}
\caption{
\label{fig:pc_val_1f}
Left: Calculation of the difference of the massive and logarithmic approximations for the inclusive $b$-jet cross section for the {\tt exp} tagging approach. The cross-section difference in the small-mass limit is fitted and also compared to a direct calculation. Right: Calculation of the difference of the massive and logarithmic approximations for the inclusive $b$-jet cross section for the {\tt exp(thr)} tagging approach.
}
\end{center}
\end{figure}

The numerical study of power corrections presented in the right plot of \fig{fig:pc} has been performed with a private implementation of the massive calculation for the process $pp \to b \bar b +X$ at NLO QCD, as well as the massless limit of the calculation as documented above.
As a validation of the implementation and the procedure discussed above, we calculate the left-hand side of \eqn{eq:4fsmL} for decreasing values of $m_b$ and study the behaviour of the cross section in this limit.
Specifically, we consider the first bin of the inclusive $p_{T,b\text{-jet}}$ distribution in \fig{fig:pc} (i.e.\  the inclusive $b$-jet cross section with $p_{T,b\text{-jet}}\in[40,60]$\,GeV and $|\eta_{b\text{-jet}}| < 2.5$), where the power corrections in $m_b$ are largest,  in \fig{fig:pc_val_1f}.
In the left plot we show the calculation based on the {\tt exp} tagging procedure, while in the right plot the result for the {\tt exp(thr)} tagging procedure is shown.
For the {\tt exp} tagging procedure on the left the difference of the massive and logarithmic calculations approaches a constant behaviour in the $m_b\to0$ limit, indicating that ${\rm d}\sigma_{L}$ exactly cancels  the logarithmic $m_b$ contributions.
Furthermore, the fitted ${\rm d}\sigma_{0}$ result in the $m_b\to0$ limit is in excellent agreement with its direct calculation.
For the {\tt exp(thr)} tagging procedure in the right plot the bottom quark (which tags the jet) is required to have a minimum $p_{T,b} = 5$\,GeV.
Also in this case, the cross-section difference approaches a clear plateau in the $m_b\to 0$ limit, 
which can be used to fit the constant contribution ${\rm d}\sigma_{0}$,
since it is not directly calculated for the {\tt exp(thr)} tagging, as discussed before.
We note that for this selection, we did not find any significant difference between the two tagging procedures.

\begin{figure}[t!]
\begin{center}
\begin{tabular}{cc}
\includegraphics[width=.42\textwidth]
{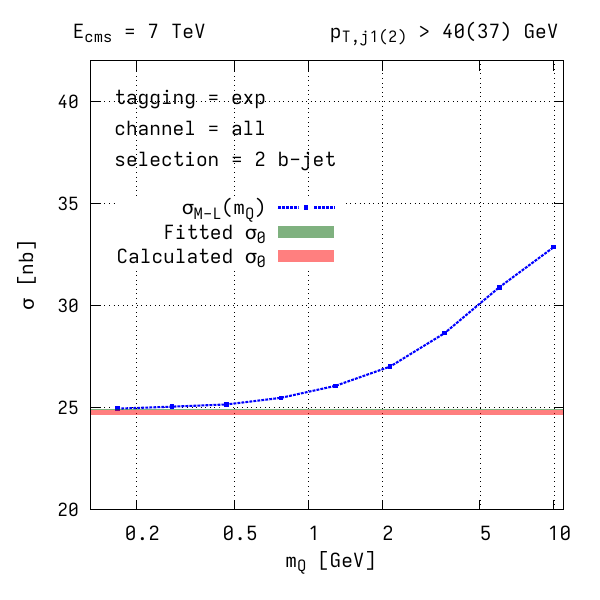}
&
\hspace{0.25cm}
\includegraphics[width=.42\textwidth]
{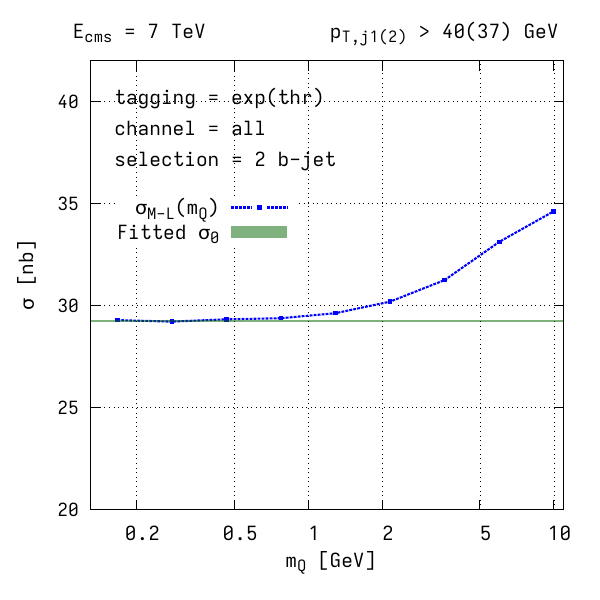}
\\
\end{tabular}
\vspace*{1ex}
\caption{
\label{fig:pc_val_2f}
As in \fig{fig:pc_val_1f}, now for a final-state selection requiring two $b$-jets and with asymmetric cuts of $p_{T,j} > 40(37)$\,~GeV on the (sub)leading jet.}
\end{center}
\end{figure}

Having correctly determined  both ${\rm d}\sigma_{L}$ and ${\rm d}\sigma_{0}$
(and validated the procedure to extract them),
we obtain the power corrections in $m_b$ using \eqn{eq:PC}. Specifically, we repeated this approach for every bin in the $p_{T,b\text{-jet}}$ spectrum in the right plot of \fig{fig:pc}. 

As a further validation of the logarithmic cross section for the {\tt exp(thr)} tagging procedure, we repeat the analysis with a selection requiring two $b$-jets. Both $b$-jets are required to satisfy $|\eta_{b\text{-jet}}|<2.5$ and $p_{T,b\text{-jet}}\gtrsim37$\,GeV, with the additional requirement $p_{T,b\text{-jet}_1}\gtrsim40$\,GeV on the leading $b$-jet. Requiring two $b$-jets eliminates the (pseudo-)collinear enhancements associated with initial- or final-state $g\to b\bar b$ splittings and thereby increases the relative sensitivity to the collinear enhancement induced by the minimum $p_{T,b}$ requirement. The results are shown in \fig{fig:pc_val_2f}. For both the {\tt exp} (left) and {\tt exp(thr)} (right) tagging procedures, the difference between the massive and logarithmic calculations approaches a constant in the $m_b\to0$ limit. For {\tt exp} tagging, the extracted constant is also in agreement with the direct calculation.

\setlength{\bibsep}{3.1pt}
\renewcommand{\em}{}
\bibliographystyle{apsrev4-1}
\bibliography{MiNNLO}

\clearpage

\end{document}